\documentclass{article}
\usepackage[numbers]{natbib}
\usepackage[colorlinks,citecolor=blue,urlcolor=blue,filecolor=blue]{hyperref}
\usepackage{fullpage}
\usepackage{amssymb}
\usepackage{amsmath}
\usepackage{amsfonts}
\usepackage{graphicx}
\usepackage{caption}
\newtheorem{theorem}{Theorem}

\newtheorem{definition}[theorem]{Definition}
\newtheorem{example}{Example}

\def\condindep{\perp\!\!\!\perp}

\usepackage{tikz}
\usetikzlibrary{calc,arrows.meta}

\begin{document}

\title{A Causal Analysis of the Plots of Intelligent Adversaries}
\author{Preetha Ramiah\footnote{Department of Statistics, University of Warwick, UK.}, David I. Hastie\footnote{School of Mathematical Sciences, Queen Mary University of London, UK.}, F.O. Bunnin\footnote{Black and Field Ltd, UK and Natwest Markets plc, UK.}, Silvia Liverani\footnote{School of Mathematical Sciences, Queen Mary University of London, UK.}, Jim Q. Smith\footnote{Department of Statistics, University of Warwick, UK and The Alan Turing Institute, UK.}}
\date{}
\maketitle

\begin{abstract}
In this paper we demonstrate a new advance in causal Bayesian graphical
modelling combined with Adversarial Risk Analysis. This research aims to support
strategic analyses of various defensive interventions to counter the
the threat arising from plots of an adversary. These plots are characterised by a sequence of preparatory phases that an adversary must necessarily pass through to achieve their hostile objective. To do this we first define a new general
class of plot models. Then we demonstrate that this is a causal graphical family of models - albeit with a hybrid semantic. We show  this continues to be so even in this adversarial setting. It follows that this causal graph can be used to guide a Bayesian decision analysis to
counter the adversary's plot. We illustrate the causal analysis of a plot with details of a decision analysis designed to frustrate the
progress of a planned terrorist attack.
\end{abstract}

\section{Introduction}

Dynamic Bayesian Networks (DBNs) \cite{Korb,Aglietti} 
 have now been established and successfully applied in many areas to support
strategic analyses of observed \emph{idle} systems - i.e. dynamical systems
not subject to control. These have then been extended to dynamic causal
models. In this way, using various plausible algorithms the predictive distributions of
idle systems can be extended to predict what might happen when that
system is controlled or - in causal terminology - subjected to various \emph{%
interventions}. General classes of graphically based causal algebras - and
not just those based on vanilla Bayesian Networks (BNs) - are now
sufficiently well defined to provide justifiable ways to apply formally this
causal reasoning and so form part of a defender's decision analysis to
support a defender in foiling the activities of an identified adversary. 

To date most graphical causal analyses of this kind have focused on domains
where the predicted impact of any interventions is on inanimate or
collaborative environments. However, in \cite{Ramiahetal23} we developed a
new Bayesian methodology that customised causal algebras to take account of
the likely potential intelligent reactions of an adversary who hears that an
intervention has taken place and who can then strive to mitigate the effects
of that intervention. We described the flexibility and potential for such
methods as part of forensic and strategic analyses where various what-if
studies can be given quantitative support within a new graphically supported
Bayesian decision analysis of an adversarial domain.

There were several challenges to be surmounted before this methodology could
be applied. In many of our targeted adversarial domains, the standard BN was
not the best graphical framework to express their idle processes. Therefore, we provided a more generic definition of a graphically based causal
analysis that was rich enough to contain classes of models typically used
for adversarial modelling. We then described how, by adapting and focusing
methodologies from Adversarial Risk Analysis (ARA) \cite{Rios09, Rios16,Banksbook, RiosInsua} we could formally embed graphical
causal models into this inferential framework.

The ARA methods proved particularly useful because - unusually for a game
theoretic methodology - they are asymmetrical. As in a Bayesian causal
analysis, they can support a single decision maker - here a defender $D$ against
their rational adversary $A$. The method assumes a Subjective Expected Utility (SEU) maximising defender $D$ is faced with a rational adversary $A$ who in \cite{Ramiahetal23} is also assumed to be an SEU maximiser. The
methodology thus accounts for $D$'s predictions of $A$'s intelligent
reactions to their interventions and provides $D$ with a methodology to
embed the adversarial nature of the domain into their SEU scores. 

This asymmetric approach allows us to embed ARA into an analysis that can be properly labeled as causal. Furthermore, for the types of strategic adversarial settings between a specific defender and their adversary, we argue that the SEU assumptions of the adversary are
plausible ones \cite{Ramiahetal23}, unlike in more economic applications of game theory. We have called this class of causal models \emph{intelligent}.

We argued in \cite{Ramiahetal23} that, if a defender truly believes that their chosen graphical framework is causal, this must entail 
that the structural information
encoded in the graph is not only invariant to various types of intervention $D$ might choose, but also represent other intelligent people's structural
beliefs - here specifically assumed to be the adversary $A$. So, within this
causal setting it is plausible for $D$ to assume that their framing graph $%
\mathcal{G}$ is common knowledge with $A$ \cite{Allpranisgame}. This
additional strong assumption, not common in current ARA applications, facilitates $D$ putting themselves in $A$'s shoes, not only in terms of $A$'s utility, but also their structural beliefs. Therefore, it enhances $D$'s ability to predict $A$'s responses to each putative intervention $D$ might
make. In particular, it is then possible for $D$ to decompose their model.
We demonstrate below these predictions reduce into $D$ needing to double guess a small
number of $A$'s conditional densities about various features of the problem
rather than $A$'s full density. The challenges in $D$'s modelling are thus
dramatically simplified. This makes this class of models not only plausibly
scalable, but also transparent in the sense that the rationale behind $D$'s
analysis is supported by this common knowledge graph.

In \cite{Ramiahetal23} we developed this formal methodology. However, we
confined our illustrations to rather trivial domains where the efficacy of
the new technology was self-apparent. In this paper we demonstrate how this
whole causal graphical modelling methodology could fulfill the promise of
actually being capable of scaling up to address strategic analyses and real
time decision support within the complex adversarial settings that might be
encountered by a defender. Here we apply the new graphically framed causal methodologies to a challenging complex class of dynamic model we call \emph{plot models} \cite{BuuninandSmith19,ShneivB&S21} that are now used to model several different adversarial domains.

We define a plot in an adversarial setting to be a mission planned by an adversary where, to reach a successful conclusion, the agent of that adversary must first complete a sequence of preparatory \emph{phases}. Although such phases will typically be hidden from the defender $D$, both $D$ and $A$ will be aware of the nature of the phases of the attack $A$ must accomplish and also the available current technologies needed for $A$ to pass from one preparatory phase of an attack to the next. Using our running example of a terrorist bombing plot we demonstrate how to use intelligent causal graphs to support $D$'s forensic, strategic and real time decisions when both $D$ and $A$ believe a plot model defines the idle - i.e.
the uncontrolled - processes.

In Section \ref{sec:gcap} We briefly review this general inferential framework of Bayesian causal graphical models. Then, in Section 3, we prove that the
class of idle plot models is graphical in the causal sense we defined in \cite{Ramiahetal23}. We model plots
with a novel class  $\mathbb{G}$ of causal graphical model - a hybrid class
of more well studied causal graphs. In Section 4 we then adapt ARA
methodologies so that they apply to this particular causal graphical
framework. 

This class is then used to model intelligent cause applied to plots whose structure is expressed by a common knowledge graph. We demonstrate how a defender might model probabilistically a particular plot, including an adversary's intelligent reactions to various types of defensive intervention. We proceed to explore the use of the graphical framework to study the impacts of intelligent causal algebras that can be designed for the plot models, illustrating these by a running example of a terrorist bombing plot. We end the paper with a short discussion about how these methods can be further extended so that these apply to wider classes
of plots.

\section{General Causal Algebras and Plots}
\label{sec:gcap}


Within a Bayesian decision analysis, assume an SEU maximising defender $D$ has utility function $U_{D}(\boldsymbol{e})$
on a vector of attributes/\emph{effects} $\boldsymbol{e}$ and \emph{%
interventions} $d\in \mathbb{D}$. Henceforth $\mathbb{D}$ includes
the option of doing nothing, denoted by $d_{\emptyset }$. Here this vector $%
\boldsymbol{e}$ is assumed to be a sub-vector of the random vector $%
\boldsymbol{\xi }$ that $D$ uses to describe all relevant features of the
idle and intervened processes. The factorisations of the densities $\left\{
p_{d}(\boldsymbol{\xi })\in \mathbb{P}_{\mathcal{G}}:d\in \mathbb{D}\right\} 
$ of $\boldsymbol{\xi }$ are assumed by $D$ to respect the (possibly
coloured) graph $\mathcal{G}\in\mathbb{G}$ where $\mathbb{G}$ is a class of
such models with a particular semantic $\mathcal{S}_{\mathbb{G}}$ relating
the topology of $\mathcal{G}$ to a particular factorisation of a density. Examples of these such classes $\mathbb{G}$ could be a BN \cite{Pearl2000,Korb}; a factor graph \cite{logfactorgraph}; event based graphs such as state transition diagrams or chain event graphs; or hybrids of all of these - as illustrated here by plot models.

To perform an SEU analysis $D$ calculates their SEU scores
\begin{equation}
\overline{U}_{D}(d)\triangleq \int U_{D}(\boldsymbol{e})p_{d}(\boldsymbol{e}%
)d\boldsymbol{e}  \label{SEU defender}
\end{equation}%
where $p_{d}(\boldsymbol{e})$, $d\in \mathbb{D}$, denotes the density of the
effects $\boldsymbol{e}$ marginalised from the full joint density $p_{d}(%
\boldsymbol{\xi })$, $d\in \mathbb{D}$, whose structure is represented by a
graph $\mathcal{G}\in \mathbb{G}$ chosen from an appropriate class. In the
case when a defender faces a plot, its success will typically be defined as a
function of how many phases $A$ manages to complete before they are caught.
Given $U_{D}(\boldsymbol{e}),$ $D$ now needs a plausible transformation $\Phi $ which
maps a model described by a graph $\mathcal{G}$ and its customised semantics 
$\mathcal{S}_{\mathbb{G}}$ of the density of the shared idle process $p(%
\boldsymbol{\xi })\triangleq p_{d_{\emptyset }}(\boldsymbol{\xi })$ into a
class of densities $\left\{ p_{d}(\boldsymbol{\xi })\in \mathbb{P}_{\mathcal{%
G}}:d\in \mathbb{D}\right\} $ - describing what happens to a plot after $D$
applies each control $d \in \mathbb{D}\setminus \left\{d_{\emptyset }\right\}$.

For simplicity, in our running example in this paper we will assume that the defender's utility function $U_{D}(\boldsymbol{e})$ takes the form 
\begin{equation*}
U_{D}(\boldsymbol{e})=\left\{ 
\begin{array}{cc}
1 & \text{plot foiled and agent disabled} \\ 
u_{D} & \text{plot foiled but agent remains free} \\ 
0 & \text{plot succeeds}%
\end{array}%
\right.
\end{equation*}%
where $0 < u_{D} < 1$.

Once $D$ specifies such a map $\Phi $ they will be able to calculate the
scores $\left\{ \overline{U}_{D}(d):d\in \mathbb{D}\right\} $ required to
provide a Bayesian decision support for a strategic or real time decision
analysis. Such a map $\Phi $ could be completely general. However, in causal
settings like the one we describe below, there are various practical and
technical reasons why a \emph{graphical} causal analysis has proved useful \cite{Robins86,Pearl2000,Spirtesetal1993,Shafer}.
Perhaps the most important reason is that an appropriately customised class $\mathbb{G}$ of graphs can be used systematically to embed into an analysis
various descriptions of the structural and quantitative expert judgements that are intrinsic to the domain of application - here the progress of a plot.

If the class $\mathbb{G}$ is chosen wisely, the appropriately selected or
elicited (possibly coloured) graph $\mathcal{G}\in\mathbb{G}$ then provides
a non-technical \emph{interface} between how $D$ might describe the
relationships between causally associated extracted features of the idle and
the intervened system. Crucially, the graph $\mathcal{G}$ also provides a
formal framework around which to construct families of probability models
consistent with these extracted explanations of how the processes might
unfold. The semantics $\mathcal{S}_{\mathbb{G}}$ of $\mathbb{G}$ enables the
topology of $\mathcal{G}$ to frame the \emph{quantified} predictive model of
what might happen to the system when a defender's intervention is applied. 

More explicitly, $\mathcal{G}$ and $\mathcal{S}_{\mathbb{G}}$ determines the 
\emph{factors} 
\begin{equation}
\mathcal{P}_{d\mathcal{G}}\triangleq \left\{ p_{d1\mathcal{G}}(\boldsymbol{%
\xi }),p_{d2\mathcal{G}}(\boldsymbol{\xi }),\ldots ,p_{dm\mathcal{G}}(%
\boldsymbol{\xi }):d\in \mathbb{D}\right\}   \label{factors}
\end{equation}%
the decision analyst will need to elicit from $D$ to construct the margins $%
\left\{ p_{d}(\boldsymbol{e}):d\in \mathbb{D}\right\} $. These can then be
used to calculate their SEU scores. The semantics $\mathcal{S}_{\mathbb{G}}$
of $\mathcal{G}$ will also determine the \emph{form of the (shared) function}
$f_{\mathcal{G}}$: a formula which composes the factors in $\mathcal{P}_{d%
\mathcal{G}}$ that determine  the joint probability densities $\left\{ p_{d%
\mathcal{G}}(\boldsymbol{\xi }):d\in \mathbb{D}\right\} $, where
\begin{equation}
p_{d\mathcal{G}}(\boldsymbol{\xi })=f_{\mathcal{G}}\left( p_{d1\mathcal{G}}(%
\boldsymbol{\xi }),p_{d2\mathcal{G}}(\boldsymbol{\xi }),\ldots ,p_{dm%
\mathcal{G}}(\boldsymbol{\xi })\right) \in \mathbb{P}_{\mathcal{G}}
\label{prob zi}
\end{equation}%
is needed to calculate expectations of these utilities. 

In the next section, for the first time, we apply this graphically based
construction to the class of plot models defined below. Examples of how
these are defined for common causal graphical models such as the BN and the
Chain Event Graph (CEG), and how these satisfy Definition 1 below are given in \cite{Ramiahetal23}.

A causal algebra \cite{PetersRSS} is based on a premise that it will provide a
commonly accepted description of the \textit{causal} processes to be driving both
the idle and any intervened system. So a central assumption of such a causal
analysis is that the topology of $\mathcal{G}$ is \emph{invariant} to the
class of interventions $d\in \mathbb{D}$ and accurately represents the
structural information embedded in the \emph{reasoning of rational modelers}. However, once $\left( \mathcal{G},\mathbb{G}\right) $ is in place, $D$ can plausibly assume the following statements when $A$ is rational and $\mathcal{G}$ is causal. 

\begin{enumerate}
\item The structural framework $\mathcal{G}$ holds true for all
interventions $d\in \mathbb{D}$ contemplated by $D$.

\item The \emph{structural information} is common knowledge between $D$ and $A$: $D$ believes $\mathcal{G}$ and the adversary $A$ also believes $\mathcal{G}$.

\item Many factors - including those associated with the underlying science
and generally accepted theory used by $D$ to embellish $\mathcal{G}$ into
their full probability model - will be shared by $A$ (i.e. be common knowledge with them) and be invariant to $D$'s choice of $d\in \mathbb{D}$.
\end{enumerate}
Note that our applications are ones where $A$ has similar capabilities to $D$. 

Here we recall the definition of a general causal map:
\begin{definition}
Call a map $\Phi$ \emph{causal} with respect to a set of
interventions $\mathbb{D}$ if, for every $d\in \mathbb{D}$, there is a
deterministic well defined map $\Phi $ such that 
\begin{eqnarray*}
\Phi &:&\mathbb{P}_{\mathcal{G}}\times \left\{ \mathcal{P}_{d\mathcal{G}%
}^{1}:d\in \mathbb{D}\right\} \rightarrow \mathbb{P}_{\mathcal{G}} \\
&:&\left( p_{\mathcal{G}}(\boldsymbol{\xi }),\left( p_{d1\mathcal{G}}(%
\boldsymbol{\xi }),p_{d2\mathcal{G}}(\boldsymbol{\xi }),\ldots ,p_{dm%
\mathcal{G}}(\boldsymbol{\xi })\right) \right) \mapsto p_{d\mathcal{G}}(%
\boldsymbol{\xi })
\end{eqnarray*}%
where $p_{d\mathcal{G}}(\boldsymbol{\xi })$ is defined by Equation (\ref{prob zi}). The
formulae $\left\{ f_{\mathcal{G}}\left( p_{d1\mathcal{G}}(\boldsymbol{\xi }), p_{d2\mathcal{G}}(\boldsymbol{\xi }), \ldots, p_{dm\mathcal{G}}(\boldsymbol{\xi })\right): d\in \mathbb{D}\right\} $ together constitute the \emph{%
causal algebra} of $\Phi $.
\end{definition}

Our first challenge is to define a class of models $\mathbb{G}$ with a
semantic $\mathcal{S}_{\mathbb{G}}$ suitable for describing the critical
structure that are typical of plots. This class must have the property that $\mathcal{%
G\in }$ $\mathbb{G}$ can give rise to a graph which is causal in the sense given above for common classes of interventions $d\in \mathbb{D}$ that $D$ might
contemplate making. Certain idle plot models, similar to the terrorist plot described in Example 1 of Section \ref{subsec:intro}, have already appeared in the literature \cite%
{BuuninandSmith19,ShneivB&S21}. In the next section we demonstrate how the idle systems of more general classes of plot can also be
represented. We then show how the new ARA technologies can be adapted as described in \cite{Ramiahetal23} to take account of adversary $A$'s reactions to any interventions that can be plausibly and transparently modelled by the type of causal maps discussed above for a general class of such plots.

\section{Graphs and Inference for Idle Plot Models}

\subsection{Introduction}
\label{subsec:intro}
Plot models are defined here as ones whose idle system is a particular class
of Bayesian dynamic three level hierarchical Markov model.

\begin{enumerate}
\item The deepest level of this hierarchical model is a latent discrete
times series $\mathbf{W}^{T}\triangleq \left\{ W_{t};t=1,2,\ldots,T\right\}$.
Its disjoint states $\left\{ w_{0},w_{1},w_{2},\ldots ,w_{m}\right\} $
denote the preparatory state of a plot the agent has at time $t=1,2,\ldots,T$.
The state $w_{0}$ - called the \emph{inactive} state - is absorbing and
denotes that the plotter is no longer pursuing the current plot. Call all
other states \emph{active} states.

\item Conditional on $A$'s agent being in a particular active state of a
plot, there are certain tasks that the agent needs to complete before they are
ready to proceed to a subsequent state. At each time, the intermediate layer
of the hierarchical model is a \emph{task vector} $\boldsymbol{\theta }%
_{t}=\left( \theta _{t1},\theta _{t2},\ldots ,\theta _{tn}\right) $
consisting of component tasks - often indicator variables. Sub-vectors $%
\boldsymbol{\theta }_{tI(w_{j})}$ of the task vector $\boldsymbol{\theta }%
_{t}$ will be associated with each active state $w_{j}$, $j=1,2,\ldots ,m$. Let $\boldsymbol{\theta }^{T}\triangleq \left\{ \boldsymbol{\theta }%
_{t}:t=1,2,\ldots ,T\right\}$.

\item The surface level of the hierarchy is $\boldsymbol{Z}^{T}\triangleq \left\{ \boldsymbol{Z%
}_{t}:t=1,2,\ldots ,T\right\}$,  a time series of observations called \emph{intensities}. These are functions of information
routinely available to $D$. We assume here that these functions have already
have been elicited from experts or formally extracted from data available to
police. Each component $\boldsymbol{Z}_{ti}$ of the intensity vector will be
chosen so that they are particularly indicative of whether or not an agent
is engaging in a particular task $\theta _{ti}$, with $i=1,2,\ldots ,n$ at time $t=1,2,\ldots,T$.
\end{enumerate}

Note that any plot therefore has three different types of structural
information that characterises it, all of which can be expressed in natural
language. The first are the phases of the plot and how it is possible to
transition between these. The second are the tasks (or omission of tasks)
that characterise activities that need to be enacted during a particular
phase of a plot. The third level models how someone's engagements in tasks
and their manifestation in what police might be able to observe depend upon
each other and across time, both for an inactive person and a person at each
of the phases of the plot.

\subsubsection{Notation}
In the remainder of the paper we use $t$ as the time index, $m$ for the number of distinct \textit{active} phases, and $n$ for the number of tasks and thus the number of task intensities; a subscript refers to a single indexed variable, while a $t$ superscript refers to all time indexed variables up to and including $t$. $\mathcal{G}^{S}$ is a graph of a 2 Time-slice Dynamic Bayes Net (2TDBN) with tasks and intensities as the vertex set, $\mathcal{G}^{B}$
is a bipartite graph with active states and tasks as the vertex set, and $\mathcal{G}^{W}$ is a graph with the active states as the vertex set i.e., the sequential phases of the plot. The hybrid graph consisting of the concatenation of these three graphs provides the structure of the idle plot model onto which the causal algebra is applied to facilitate adversarial decision analysis.

We now illustrate with an example how this structural information can be depicted and then embellished into a full dynamic Bayesian model.

\begin{example}\label{ex1}
Police have learned that an adversary $A$ has persuaded an agent, known to $D$, to perpetrate a lone suicide bombing attack on the general public. To do this the agent must complete a number of
preparatory phases of activity. He must first learn how to make a bomb - a preparatory step we denote by state $w_{1}$. Associated tasks in the
middle layer of the hierarchy might be to meet and learn this skill from an accomplice who already knows how to make a bomb or alternatively to travel
abroad to receive the necessary training - so engaging in buying tickets to an appropriate destination or finally and less reliably, to learn from the
dark web how to make a bomb. Denote whether the agent is engaging in these three tasks at time $t$ by the sub-vector $\boldsymbol{\theta }_{tI(w_{1})}$.
The agent is under surveillance by $D$ who extracts often routinely collected electronic signals that hint that these tasks might be underway. We
call such tell-tale signs intensities which are represented by
certain sub-vectors of components $\boldsymbol{Z}_{I(w_{1})}^{T}$ of $%
\boldsymbol{Z}^{T}$.

Once trained the agent will then need to source the necessary bomb
making materials (state $w_{2}$). This phase involves locating and then
physically obtaining all its ingredients and components. These tasks could
be achieved either by the agent himself sourcing materials or by these being
delivered directly to $A$, denoted by the sub-vector $\boldsymbol{\theta }%
_{tI(w_{2})}$. Again, such activities could be suggested, for example, by
routine surveillance, phone tapping and checks of sales by suppliers of
potentially hazardous chemicals, $\boldsymbol{Z}_{I(w_{2})}^{T}$.

Before or after this activity the agent will also need to identify and reconnoitre a target  (state $w_{3}$) to determine where he could cause the
most drama and spread the most fear, individually or working with
accomplices of the agent. The tasks associated with this phase could include
investigating the demographics and defences of different candidate sites, which could involve visiting potential target sites and/or electronically
exploring maps of potential target areas. These two possible tasks are represented in $\boldsymbol{\theta }_{tI(w3)}$. Again such activities might
be captured by police monitoring the web, their interception of meta data
associated with phone messages, CCTV and direct observation, $\boldsymbol{Z}%
_{I(w_{3})}^{T}.$

Finally the agent will need to arm himself with the bomb and travel to the target armed with the bomb before detonating it. The state $w_{4}$ has
two component tasks $\boldsymbol{\theta }_{tI(w_{4})}$ of arming himself and
travelling to the scene, with related activities captured by $\boldsymbol{Z}_{I(w_{4})}^{T}$.
\end{example}

\subsection{A Generic Class of Plot Models}
\label{subsec:plot_specification}

\subsubsection{An Illustration of the Components of a Plot Model}
In this paper we demonstrate that idle models of plots like the one in Example \ref{ex1}
are indeed \emph{graphical}. It will then
follow that they can be endowed with an associated graphical causal algebra.
This can then be used - within this adversarial setting - to guide police in
their countermeasures against an adversary. We show here that the most
appropriate causal graphical class for representing a plot - able to capture
all three qualitative features mentioned above - is one which is a hybrid of
more familiar dynamic graphs.

Henceforth, assume $D$'s contemplated interventions $d\in \mathbb{D}$ -
designed to frustrate the progress of the plot - will be planned to be
enacted at a time denoted by $t_{0}$. The effect of such interventions will
therefore be felt through times $t=t_{0},\ldots ,T$. Let the filtration $%
\mathcal{F}_{t_{0}}$ denote all information available to $D$ at time $t_{0}$
- both routinely collected and via incidental intelligence. Without loss of
generality and for transparency, the conditioning on $\mathcal{F}_{t_{0}}$
will henceforth remain implicit in the notation we give below.

For $t=1,2,\ldots ,T$, write $\boldsymbol{\xi }_{t}\triangleq (\boldsymbol{w}%
_{t}^{T},\boldsymbol{\theta }_{t}^{T},\boldsymbol{z}_{t}^{T})$ where 
\begin{equation*}
\boldsymbol{w}_{t}^{T}\triangleq \left( w_{t},w_{t+1},\ldots, w_{T}\right) ,%
\boldsymbol{\theta }_{t}^{T}\triangleq \left( \boldsymbol{\theta }_{t},%
\boldsymbol{\theta }_{t+1},\ldots, \boldsymbol{\theta }_{T}\right) ,%
\boldsymbol{z}_{t}^{T}\triangleq \left( \boldsymbol{z}_{t},\boldsymbol{z}%
_{t+1},\ldots, \boldsymbol{z}_{T}\right)
\end{equation*}%
are the phases, tasks and intensities defined above. To illustrate each of these components consider a simple model of the bomb plot above where there are seven states $\left(
w_{0},w_{1},w_{2},w_{2\star},w_{3},w_{3\star},w_{4}\right) $ with:
\begin{table}[ht]
\centering
\begin{tabular}{ll}
$w_{0}$& agent not actively engaged in bombing plot\\
$w_{1}$& agent learning how to make a bomb\\
$w_{2}$& agent acquiring bomb making materials (target not yet identified)\\
$w_{2\star}$& agent, who has already identified a target, acquiring bomb making materials\\
$w_{3}$& agent identifying and reconnoitring a target (bomb making materials not yet acquired)\\
$w_{3\star}$& agent, who had already acquired bomb making materials, identifying and reconnoitring a target\\
$w_{4}$& agent armed and travelling to the target to detonate the bomb.
\end{tabular}
\end{table}

The binary vector of tasks $\boldsymbol{\theta }=\left( \theta_{1},\theta _{2},\theta _{3},\theta _{4},\theta _{5},\theta _{6},\theta_{7},\theta _{8}\right) $ as identified above would be:
\begin{table}[ht]
\centering
\begin{tabular}{ll}
$\theta _{1}$& extended visits from an accomplice suspected of
having bomb making skills\\
$\theta _{2}$& agent searching the internet and the dark web for bomb making instructions\\
$\theta _{3}$& agent travels abroad for bomb making training\\
$\theta _{4}$& agent directly searching for, finding and buying materials that could make a bomb\\
$\theta _{5}$& agent receives materials that could make a bomb delivered by $A$ to the agent\\
$\theta _{6}$& agent researching potential target on the web\\
$\theta _{7}$& agent constructs and arms bomb\\
$\theta _{8}$& agent physically travels to the identified target.
\end{tabular}
\end{table}

From the descriptions above we can now identify the sub-vectors associated
with the active states. They are conjectured to be
\begin{equation*} 
\begin{split}
\boldsymbol{\theta }_{I(w_{1})} &=\left( \theta _{1},\theta _{2},\theta
_{3}\right) \\
\boldsymbol{\theta }_{I(w_{2})} = \boldsymbol{\theta }_{I(w_{2\star})}&=\left( \theta _{4},\theta
_{5}\right)\\
\boldsymbol{\theta }_{I(w_{3})}=\boldsymbol{\theta }_{I(w_{3\star})}&=\left( \theta _{6},\theta
_{8}\right)\\
\boldsymbol{\theta }_{I(w_{4})}&=\left( \theta _{7},\theta
_{8}\right)
\end{split}
\end{equation*}
We have found that for many real plots the same task might be
associated with different phases of a plot. For example, here the
component $\theta_{8}$ lies in both $\boldsymbol{\theta }_{I(w_{3})}$ and $%
\boldsymbol{\theta}_{I(w_{4})}$.

The intensities $\boldsymbol{z}_{t}$ would then be associated with tell-tale signs about these tasks that are routinely available to the
defenders. Such measures will be dynamic and highly dependent on the resources of the defender and the modus operandi (MO) of the adversary and their agent.
They will also often be operationally sensitive and so we have not specified them explicitly here.

Note that activities associated with both the tasks and their intensities will
typically be dependent on each other, especially when the suspect is
engaging in a particular phase of the attack. For example in $w_{2}$ above,
the suspect will only need to have completed one of $\theta _{4}$ or $\theta
_{5}$. So, these activities are mutually exclusive and their indicators are not
independent of each other. For similar reasons we might well expect that
when a suspect is engaged in phase $w_{1}$ then typically the component
tasks $\left( \theta _{1},\theta _{2},\theta _{3}\right) $ might be
negatively dependent. Moreover, certain tasks may need to happen before
others in the task vector. Henceforth assume that there is an ordering of
the tasks such that, whatever phase the agent is in, the indexing of the
tasks respect this order.

\subsubsection{How Effects Relate to the Graph of a Causal Plot model}

For a causal analysis, $D$ will need to predict the effects $\boldsymbol{e}$
of a plot. These will necessarily be a function of the variables $\boldsymbol{\xi }_{t_{0}}$ that occur after the time $t_{0}$ of any contemplated intervention. The effects $\boldsymbol{e}$ of many plot models
that are of interest to $D$ will be a function of $\boldsymbol{w}_{1}^{T}$ alone. Indeed, for the bombing plot described above there is a single effect -
measuring the threat to public safety - which is a function of whether the
agent reaches the terminal phase $w_{4}$ or retreats into state $w_{0}$. So
inference in plot models tends to focus on the impacts felt by decisions on
this margin - the density $p_{t_{0}}(\boldsymbol{e})$ conditional on $%
\mathcal{F}_{t_{0}}$ will be a function only of the margin $p_{t_{0}}(%
\boldsymbol{w}_{1}^{T})$. For this reason, in plot models, posterior
inferences about idle systems only need to describe accurately the time course
of $\boldsymbol{w}_{1}^{T}$ posterior to $\mathcal{F}_{t_{0}}$.

We will show that the hybrid semantics of $\mathcal{G}$ - when embellished
into a new graph $\mathcal{G}^{+\text{ }}$to capture new features not needed
for the idle system, but explaining how $A$ might respond to $D$'s
interventions - can be justifiably identified as causal in the theoretical
sense described above. In particular the topology of $\mathcal{G}^{+\text{ }}
$ will be invariant to different interventions $d\in \mathbb{D}$, that $D$ might
choose. We will show that this is true even for ones that might provoke a
reaction from $A$ which enables $D$ to produce the SEU
scores they need for a decision analysis. We first describe more explicitly
how this hybrid structural graph $\mathcal{G}$ of the idle progress of a
plot is constructed.

\subsubsection{A Graphical Model to Describe Task and Intensity Relationships}

We begin by defining a new general class of models we
call plot models that is sufficiently general to embrace all the applications we have needed recently and beyond those described in \cite%
{BuuninandSmith19}. Here we propose the relationships between the
components of the tasks and intensities of a plot by the now well
established graphical model - a 2TDBN.
Assume at each time slice of this 2TDBN, $\mathcal{G}^{S}$ has
vertices $\left\{ \theta _{t1},\theta _{t2},\ldots ,\theta _{tn}\right\} $
depicting the components of the task vector $\boldsymbol{\theta }%
_{t}=\left( \theta _{t1},\theta _{t2},\ldots ,\theta _{tn}\right)$ on each time slice $t \leq T$. The probabilities associated with these
binary variables are significantly increased (or decreased) for an adversary in least one
of the active phases. This enables $D$ to discriminate between the different
phases of the plot from the noisy signals they receive through the
intensities they observe. First we define $\boldsymbol{W}^{t},\boldsymbol{\theta }^{t},\boldsymbol{Z}^{t}$ as
\begin{equation*}
\boldsymbol{W}^{t}\triangleq \left( w_{1},w_{2},\ldots, w_{t}\right) ,%
\boldsymbol{\theta }^{t}\triangleq \left( \boldsymbol{\theta }_{1},%
\boldsymbol{\theta }_{2},\ldots, \boldsymbol{\theta }_{t}\right) ,%
\boldsymbol{Z}^{t}\triangleq \left( \boldsymbol{Z}_{1},\boldsymbol{Z}%
_{2},\ldots, \boldsymbol{Z}_{t}\right).
\end{equation*}%

The first assumption about a plot model, as in \cite{BuuninandSmith19}, is
to demand that 
\begin{equation}
\boldsymbol{Z}^{T}\condindep W^{T}|\boldsymbol{\theta }^{T}
\label{hierarchical ind}
\end{equation}%
This implies that $\boldsymbol{\theta }^{T}$ is defined in a sufficiently
refined way such that $D$ believes they can only learn about the phase of a
plot $A$ is engaging in through the tasks that $A$ is undertaking.

Our second assumption is that the process defining the plot is Markov,
conditional on $\boldsymbol{W}^{T}$. Technically this means that
\begin{equation}
\left( \boldsymbol{Z}_{t},\boldsymbol{\theta }_{t}\right) \condindep \left( 
\boldsymbol{Z}^{t-2},\boldsymbol{\theta }^{t-2}\right) |\boldsymbol{Z}%
_{t-1},\boldsymbol{\theta }_{t-1},\boldsymbol{W}^{t}
\label{Markov ind}
\end{equation}%
This implies that when we know which states have been accomplished to date, if we know the current tasks and intensities, all following tasks and intensities are independent of the past tasks and intensities. 
This assumption can always be ensured by choosing the latent task states to be big enough. However, from a modelling perspective, it is good practice to keep
the vector $\boldsymbol{\theta }_{t}$ as short as possible whilst still
satisfying Equation (\ref{Markov ind}) when defining a particular plot.

A third assumption is that $\boldsymbol{Z}_{tk}$ is an intensity measure
only on the task component $\theta _{tk}$, although it may also depend on the intensities at time $t$ relating to the variables indexed with $l$ such that $l \leq k$. We define the following index sets for the intensity and task vertices, respectively in Equations (\ref{Observational ind}) and (\ref{task ci}): with $S_k \subseteq \{1, \ldots, k\}$, $Q_{t}(k) \equiv S_{k-1}$ are the indices of contemporaneous parents of the variable indexed by $k$, and $R(S_k) = \{1, \ldots, n\}\setminus S_k$ for $k=1, \ldots,n$.

This third assumption can then be stated as:
\begin{equation}
\boldsymbol{Z}_{tk}\condindep \left( \boldsymbol{Z}^{t-1},\boldsymbol{\theta }%
^{t-1},\boldsymbol{Z}_{R(Q_t(k))},\left\{ \theta
_{ti}:1\leq i\neq k\leq n\right\} \right) |\theta _{tk},\boldsymbol{Z}%
_{Q_{t}(k)}  \label{Observational ind}
\end{equation}%
The selection of the appropriate functions of typically vast streams of noisy data to serve as
these intensities requires expert judgements, often together with sophisticated
signal extraction algorithms trained on the most benign scenarios. This is beyond
the scope of this paper and too sensitive to describe here. However, an
illustration of this process as it applies to plots associated with vehicle
attacks is described in \cite{BuuninandSmith19} and its supplementary
materials.

For most plots, different tasks within a single time slice will be dependent
on each other. The class of plot models expresses these dependencies as a 2TDBN. So a fourth assumption that we demand for a plot model is that $D$ believes that components $\theta _{tj}$, $%
k=1,2,\ldots ,n$ of the task vector $\boldsymbol{\theta }_{t}$ satisfy
\begin{equation}
\theta _{tk}\condindep \boldsymbol{\theta }^{t-2},\boldsymbol{\theta }%
_{R(Q_{t}(k))},\boldsymbol{\theta }_{R(Q^\prime_{t-1}(k))}|\boldsymbol{%
\theta }_{Q_{t}(k)},\boldsymbol{\theta }_{Q_{t-1}^{\prime }(k)},\left\{W_t=w_j \right\}
\label{task ci}
\end{equation}
where $Q^\prime_{t}(k) \equiv S_{k}$.
%
Note that such models contain the
distribution free versions of the Dynamic Linear Model (DLM) \cite{WestHarrison1997} where $%
Q_{t-1}(k)=\emptyset $ and the transition matrix is upper
triangular; and the Multiregression model \cite{Queen} when $%
Q_{t}(k)=\emptyset $ and $Q_{t-1}^{\prime }(k)=\left\{ k\right\} $. The plot
model described in \cite{BuuninandSmith19} assumed that $Q_{t-1}(k)=\emptyset $.

A critical implication of this assumption is that task components are \emph{homogeneously ordered}: the parents of $\theta _{tj}$ do not depend on the state $w_{j}$ at time $t$, $j=0,1,2,\ldots,m$. Formally this is not a substantive issue: a saturated graph would automatically satisfy this condition. However, from a practical point of view, models where $Q_{t}(k)$ and $Q_{t-1}^{\prime }(k)$ all have small cardinality are easiest to implement.

\begin{figure} 
    \centering
\begin{tikzpicture}[main/.style = {draw, circle}, node distance={50},line width=1pt,>=stealth,scale=1.1,minimum size=15mm] 
\node[main] (theta18) at (1,1) {$\theta_{(t-1)8}$}; 
\node[main] (theta17) at (1,3) {$\theta_{(t-1)7}$}; 
\node[main] (theta16) at (1,5) {$\theta_{(t-1)6}$}; 
\node[main] (theta15) at (1,7) {$\theta_{(t-1)5}$}; 
\node[main] (theta14) at (1,9) {$\theta_{(t-1)4}$}; 
\node[main] (theta13) at (1,11) {$\theta_{(t-1)3}$}; 
\node[main] (theta12) at (1,13) {$\theta_{(t-1)2}$}; 
\node[main] (theta11) at (1,15) {$\theta_{(t-1)1}$}; 
\node[main] (theta1) at (3,15) {$\theta_{t1}$}; 
\node[main] (theta2) at (5,13) {$\theta_{t2}$}; 
\node[main] (theta3) at (3,11) {$\theta_{t3}$}; 
\node[main] (theta4) at (5,9) {$\theta_{t4}$}; 
\node[main] (theta5) at (3,7) {$\theta_{t5}$}; 
\node[main] (theta6) at (5,5) {$\theta_{t6}$}; 
\node[main] (theta7) at (3,3) {$\theta_{t7}$}; 
\node[main] (theta8) at (5,1) {$\theta_{t8}$}; 
\node[main] (z1) at (7,15) {$z_{t1}$}; 
\node[main] (z2) at (7,13) {$z_{t2}$}; 
\node[main] (z3) at (7,11) {$z_{t3}$}; 
\node[main] (z4) at (7,9) {$z_{t4}$}; 
\node[main] (z5) at (7,7) {$z_{t5}$}; 
\node[main] (z6) at (7,5) {$z_{t6}$}; 
\node[main] (z7) at (7,3) {$z_{t7}$}; 
\node[main] (z8) at (7,1) {$z_{t8}$}; 
\draw[->] (theta17) -- (theta7);
\draw[->] (theta18) -- (theta8);
\draw[->] (theta16) -- (theta6);
\draw[->] (theta15) -- (theta5);
\draw[->] (theta14) -- (theta4);
\draw[->] (theta13) -- (theta3);
\draw[->] (theta12) -- (theta2);
\draw[->] (theta11) -- (theta1);
\draw[->] (theta1) -- (z1);
\draw[->] (theta2) -- (z2);
\draw[->] (theta3) -- (z3);
\draw[->] (theta4) -- (z4);
\draw[->] (theta5) -- (z5);
\draw[->] (theta6) -- (z6);
\draw[->] (theta7) -- (z7);
\draw[->] (theta8) -- (z8);
\draw[->] (theta1) -- (theta2);
\draw[->] (theta2) -- (theta4);
\draw[->] (theta2) -- (theta3);
\draw[->] (theta1) -- (theta3);
\draw[->] (theta4) -- (theta5);
\draw[->] (theta4) -- (theta6);
\draw[->] (theta5) -- (theta6);
\draw[->] (theta6) -- (theta7);
\draw[->] (theta6) -- (theta8);
\draw[->] (theta7) -- (theta8);
\draw [dashed] (2,-1) -- (2,17);
\node[text width=3cm] at (0,17) {Time};
\node[text width=3cm] at (1.9,17) {$t-1$};
\node[text width=3cm] at (4.1,17) {$t$};
\end{tikzpicture}

    \caption{This graph represents a special case where all $z$'s are independent of every other $z$'s in the previous time slices.}\label{fig1}
\end{figure}
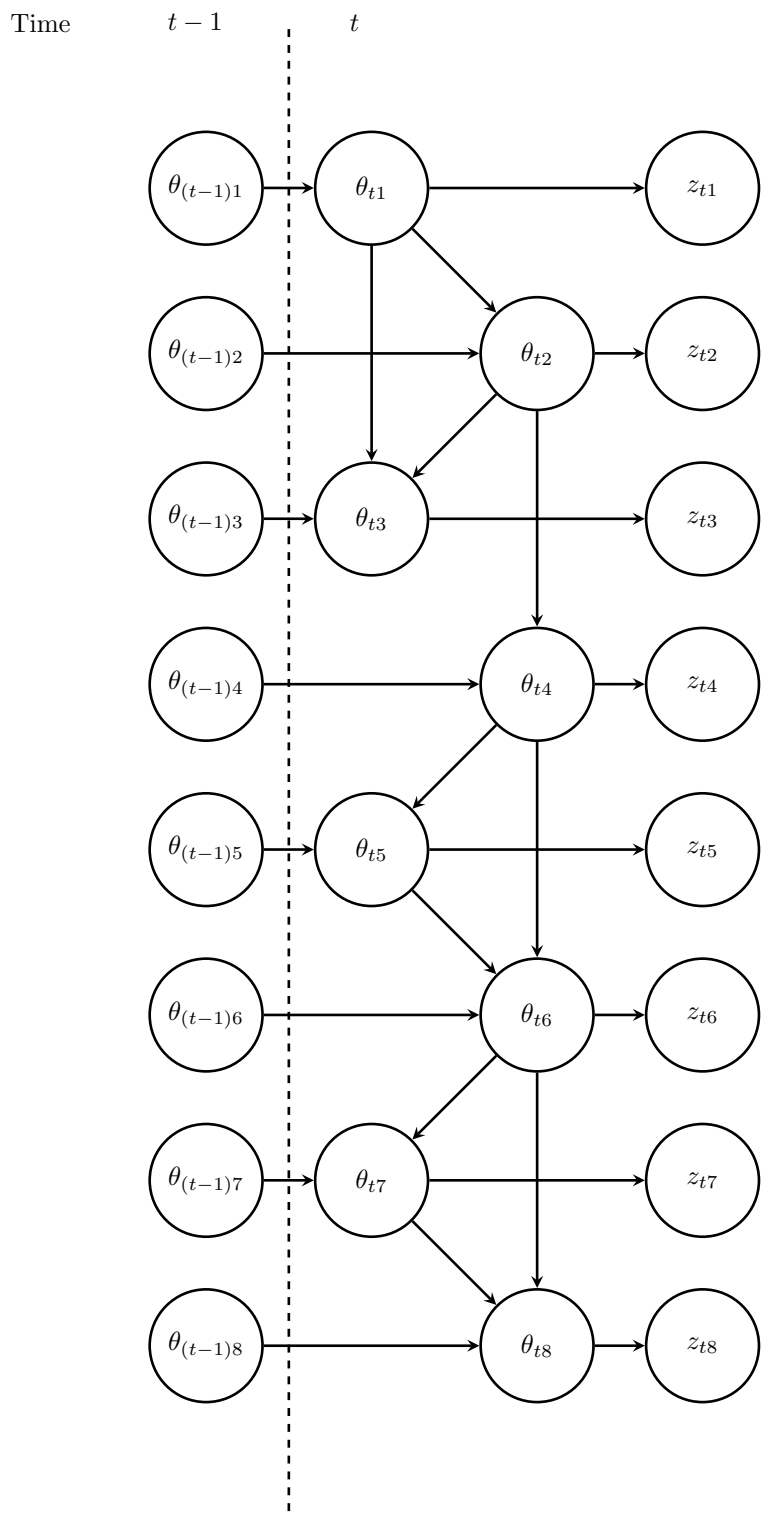

The directed graph of the 2TDBN then has the vertex set 
\begin{equation*}
V(\mathcal{G}^{S})\triangleq \left\{ \boldsymbol{Z}_{tk},\theta
_{tk}:k=1,2,\ldots,n; \quad t=1,\ldots,T\right\}
\end{equation*}%
and edge set $E(\mathcal{G}^{S})$, the directed edges from each component of $%
\boldsymbol{Z}_{Q_{t}(k)}$ into $\boldsymbol{Z}_{tk}$ and
from each component in $\boldsymbol{\theta }_{Q_{t}(k)}$ or $\boldsymbol{%
\theta }_{Q_{t-1}^{\prime }(k)}$ into $\theta _{tk}$, $k=1,2,\ldots,n.$ In our running example, one such graph is shown in Figure \ref{fig1}.

Note that all components lying in the same task set are completely connected. 
Here $\theta _{2}, \theta _{4}$ and $\theta _{6}$ are also
hypothesised as related since their probability will all be an increasing
function of that agent's expertise. Note that a change in activity - suggesting a
possible transition to a new phase - can be captured within this class
through the Markov relationships over time of each task.

Because of the Markov assumptions above and the time homogeneity of the
dependence structure of the 2TDBN, $\mathcal{G}^{S}$ can be unfolded over
all times $t=1,2,\ldots ,T$ by concatenating each of the time slice
dependencies into a single BN as described in \cite{Korb}. By unfolding $%
\mathcal{G}^{S}$ in this way and using the familiar semantics of a BN we see
that $\mathcal{G}^{S}$ implies the following factorisation of the joint density of the tasks and
intensity vectors conditional on the phases of any plot:%
\begin{equation}
p_{t_{0}}^{S}(\boldsymbol{z}_{t_{o}}^{T},\boldsymbol{\theta }_{t_{0}}^{T}|%
\boldsymbol{w}_{t_{0}}^{T})=\prod\limits_{t=t_{0}}^{T}p_{t}^{S}(\boldsymbol{%
z}_{t}|\boldsymbol{\theta }_{t},\boldsymbol{z}^{t-1},\boldsymbol{\theta }%
^{t-1})p_{t}^{B}(\boldsymbol{\theta }_{t}|\boldsymbol{z}^{t-1},\boldsymbol{%
\theta }^{t-1},w_{t_{0}})  \label{full joint obs cond}
\end{equation}%
where, for $t=t_{0},t_0 +1,\ldots ,T$. 
\begin{equation}
p_{t}^{S}(\boldsymbol{z}_{t}|\boldsymbol{\theta }_{t},\boldsymbol{z}^{t-1},%
\boldsymbol{\theta }^{t-1})=\prod\limits_{k=1}^{n}p_{tk}^{S}(\boldsymbol{z}%
_{tk}|\theta _{tk},\boldsymbol{z}_{tQ^{\prime \prime }(k)})
\label{observed densities}
\end{equation}%
and, for $j=0,1,\ldots ,m$, 
\begin{equation}
p_{t}^{B}(\boldsymbol{\theta }_{t}|\boldsymbol{z}^{t-1},\boldsymbol{\theta }%
^{t-1},w_{jt})=\prod\limits_{k=1}^{n}p_{tk}^{B}(\theta _{tk}|\boldsymbol{%
\theta }_{tQ(k),},\boldsymbol{\theta }_{(t-1)Q^{\prime }(k),}w_{jt})
\label{task densities}
\end{equation}%
where $p_{t}^{S}$, $p_{t}^{B}$ are the conditional densities of the intensities and tasks respectively. Note here that the condition that tasks are homogeneously ordered ensures
that 
\begin{eqnarray}
p_{t}^{B}(\boldsymbol{\theta }_{t}|\boldsymbol{z}^{t-1},\boldsymbol{\theta }%
^{t-1},w_{jt}) &=&p_{t}^{B}(\boldsymbol{\theta }_{t}|\boldsymbol{z}^{t-1},%
\boldsymbol{\theta }^{t-1},w_{0t})  \label{homtask factors} \\
&=&\prod\limits_{k=1}^{n}p_{tk}^{B}(\theta _{tk}|\boldsymbol{\theta }%
_{tQ(k),},\boldsymbol{\theta }_{(t-1)Q^{\prime }(k),}w_{0t})
\label{homtask factors2}
\end{eqnarray}

Let 
\begin{equation}
\mathcal{P}^{S}\triangleq \left\{ p_{tk}^{S}(\boldsymbol{z}_{tk}|\theta
_{tk},\boldsymbol{z}_{tQ^{\prime \prime }(k)}):k=1,2,\ldots
,n; \quad t=t_{0},t_{0}+1,\ldots ,T\right\}  \label{set of intensity factors}
\end{equation}%
represent the set of conditional density factors associated with the surface
observational layer of the hierarchy of the plot.

\subsubsection{The Intermediate Layer of a Plot Model: Its Graph and Factors}

Our next step is to represent graphically the critical structural information about which task engagements are indicative of which phases within a given plot. So, let $\mathcal{G}^{B}$ denote the bipartite directed graph whose vertices are $\left\{ w_{1},\ldots ,w_{m}\right\} \cup \left\{\theta _{1},\theta _{2},\ldots ,\theta _{n}\right\} $. All directed edges $e_{jk}$ will have their tail vertex, $w_{j} \in \left\{ w_{1},\ldots
,w_{m}\right\}$, the active phases of the plot and their head tasks vertex, $\theta _{k} \in \left\{ \theta _{1},\theta _{2},\ldots ,\theta
_{n}\right\}$, if and only if $\theta _{k}$ is in the task set $I(w_{j})$, $%
j=1,2,\ldots,m$, $k=1,2,\ldots ,n$.

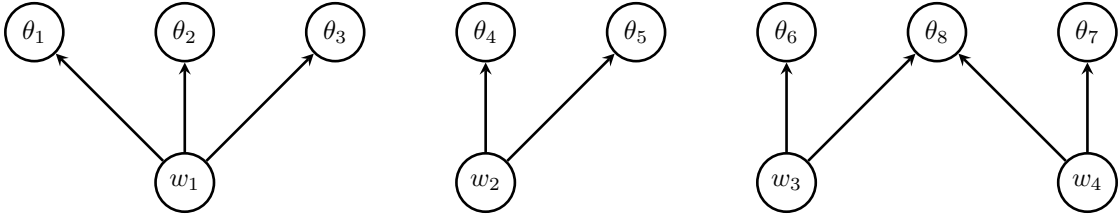
\begin{figure} 
     \centering
\begin{tikzpicture}[main/.style = {draw, circle}, node distance={20mm},line width=1pt,>=stealth] 
\node[main] (1) {$w_1$}; 
\node[main] (2) [above of=1] {$\theta_2$}; 
\node[main] (3) [left of=2] {$\theta_1$}; 
\node[main] (4) [right of=2] {$\theta_3$}; 
\draw[->] (1) -- (2);
\draw[->] (1) -- (3);
\draw[->] (1) -- (4);
\node[main] (5) [right of=4] {$\theta_4$}; 
\node[main] (6) [right of=5] {$\theta_5$}; 
\node[main] (7) [right of=6] {$\theta_6$}; 
\node[main] (8) [right of=7] {$\theta_8$}; 
\node[main] (9) [right of=8] {$\theta_7$}; 
\node[main] (10) [below of=5] {$w_2$}; 
\draw[->] (10) -- (5);
\draw[->] (10) -- (6);
\node[main] (11) [below of=7] {$w_3$}; 
\node[main] (12) [below of=9] {$w_4$}; 
\draw[->] (11) -- (7);
\draw[->] (11) -- (8);
\draw[->] (12) -- (8);
\draw[->] (12) -- (9);
\end{tikzpicture}
    \caption{The bipartite directed graph $\mathcal{G}^{B}$ for the running example. The graphs for $w_{2\star}$ and $w_{3\star}$ are the same as the graphs for $w_2$ and $w_3$, respectively.}
    \label{fig:fig GB}
    \end{figure}

So, the semantic of $\mathcal{G}^{B}$ will demand that there is a directed
edge from a phase to a task, $w_{j}\rightarrow \theta _{k}$, if and only if $D$
predicts that the probability of an agent engaging (or disengaging) in the
task $\theta _{j}$ when in phase $w_{j}$ differs from what $D$ would expect
to see when an agent is in the inactive state $w_{0}$. We adopt the
convention that the inactive state $w_{0}$ has no vertex and remains
implicit in this representation - as it does in a Reduced Dynamic Chain Event Graph (see Section \ref{ss:repp}) like the one given in Figure \ref{fig:fig3}. Denote
the sub-vector of tasks that $D$ believes $A$ might engage in or had just
stopped doing when in phase $w_{j}$, $j=1,2,\ldots ,m$ by $\boldsymbol{%
\theta }_{I(w)t}$, and its complementary vector by $\boldsymbol{\theta }%
_{I^{c}(w)t}$, $w=w_{1},w_{2},\ldots ,w_{m}$. The graph $\mathcal{G}^{B}$
therefore depicts explicitly and transparently the elicited structural
information concerning the task sets $\left\{ I(w_{j}):j=1,2,\ldots
,m\right\} $ associated with all the active phases of the plot. The graph $%
\mathcal{G}^{B}$ for our running example is given in Figure \ref{fig:fig GB}. 

This graph can then be translated into the factors of the predictive joint
densities, specifically demanding that when $i\notin I(w_{j})$ for $%
i=1,2,\ldots ,n$, $j=1,2,\ldots ,m$ and $t=1,2,\ldots ,T$%
\begin{equation}
p_{ti}^{B}(\theta _{ti}|\boldsymbol{\theta }_{tQ(i),},\boldsymbol{\theta }%
_{(t-1)Q^{\prime }(i),}w_{j})=p_{ti}^{B}(\theta _{ti}|\boldsymbol{\theta }%
_{tQ(i),},\boldsymbol{\theta }_{(t-1)Q^{\prime }(i),}w_{0})
\label{task set inclusion}
\end{equation}%
where $w_{0}$ is the inactive state. This in turn enables us to identify the
set of probability mass functions/densities associated with tasks. We need
\begin{equation}
\mathcal{P}_{0}^{B}\triangleq \left\{ p_{ti0}^{B}(\theta _{ti}|\boldsymbol{%
\theta }_{tQ(i),},\boldsymbol{\theta }_{(t-1)Q^{\prime
}(i),}w_{0}):i=1,2,\ldots ,n\right\}  \label{wo B factors}
\end{equation}%
plus those densities associated with changes from the inactive state caused
by one of the $m$ active phases of the plot, denoted for $j=1,2,\ldots, m$
here
\begin{equation}
\mathcal{P}_{j}^{B}\triangleq \left\{ p_{tij}^{B}(\theta _{ti}|\boldsymbol{%
\theta }_{tQ(i),},\boldsymbol{\theta }_{(t-1)Q^{\prime }(i),}w_{j}):i\in
I(w_{j})\right\}  \label{wi B factors}
\end{equation}%
and
\begin{equation*}
\mathcal{P}^{B}\triangleq \bigcup\limits_{j=0}^{m}\mathcal{P}_{j}^{B}.
\end{equation*}%
There are two special cases within this class which enable simple and transparent inference for a plot model. However, neither of these assumptions are needed in the
general causal extension of plots we discuss below.

Call a plot model \emph{%
regular} if for $j=1,2,\ldots ,m$ the 2TDBN can be chosen so that tasks
associated with a particular active phase can be mutually dependent.
We also call a plot \emph{naive} if when someone is in the inactive state $%
w_{0}$, for $i=1,2,\ldots ,n$ 
\begin{equation}
p_{ti0}^{B}(\theta _{ti}|\boldsymbol{\theta }_{tQ(i),},\boldsymbol{\theta }%
_{(t-1)Q^{\prime }(i),}w_{0})=p_{ti0}^{B}(\theta _{ti}|w_{0})  \label{naive}
\end{equation}%
so that all tasks under the inactive regime are independent, i.e.

\begin{equation}
p_{t}^{B}(\boldsymbol{\theta }_{t}|\boldsymbol{z}^{t-1},\boldsymbol{\theta }%
^{t-1},w_{0})=\prod\limits_{ei=1}^{n}p_{ti}^{B}(\theta _{ti}|w_{0})
\label{product ind w0}
\end{equation}%
We note that the models in \cite{BuuninandSmith19} are both regular and
naive.

Note that sampling information about the bigger set of densities in $\mathcal{P}_{0}^{B}$
 will often need to be specified by
expert judgements alone.

\subsubsection{Representing and Embellishing the Phases of a Plot}
\label{ss:repp}
Finally, at the deepest level of its hierarchy, a plot model, is a finite
discrete state space semi-Markov model. This describes the unfolding phases
of a plot that is modelled by a graphical model called a Reduced Dynamic Chain Event Graph (RDCEG). This is defined and analysed in \cite{BuuninandSmith19} and its holding time embellishments used here are in \cite{Shenvi20theory,Barrclayej}. The structural information in the RDCEG can be
represented by the coloured graph $\mathcal{G}^{W}$. The vertices of this
graph represent the underlying phases $\left\{ w_{1},w_{2},\ldots
w_{m}\right\} $ of the plot. Let $W_{t}$ be the random variable taking the
possible values of the states $\left\{ w_{1},w_{2},\ldots ,w_{m}\right\} $ at
time $t>0$ where $\left\{ w_{1},\ldots ,w_{m}\right\} $ are the vertices of $\mathcal{G}^{W}$. In RDCEG terminology these vertices are called \textit{positions} and represent active states as opposed to the inactive state $w_0$. The transitions from one
state to a different one is governed by a semi-Markov process where $w_{0}$
is assumed to be an absorbing state expressing that the agent voluntarily or
is forced to abort their mission. Note that the graph of the RDCEG $\mathcal{%
G}^{W}$ is simply a coloured version of the familiar state space diagram of
this semi-Markov process, but where we unclutter the picture by removing the
absorbing state.

\begin{figure}
    \centering
\begin{tikzpicture}[main/.style = {draw, circle}, node distance={20mm},line width=1pt,>=stealth] 
\node[main] (1) at (-2,0) {$w_1$}; 
\node[main] (2) at (0,1) {$w_2$}; 
\node[main] (21) at (2,-1) {$w_{2\star}$}; 
\node[main] (3) at (0,-1) {$w_3$};
\node[main] (31) at (2,1) {$w_{3\star}$};
\node[main] (4) at (4,0) {$w_4$};
\draw[->] (1) -- (2);
\draw[->] (1) -- (3);
\draw[->] (2) -- (31);
\draw[->] (3) -- (21);
\draw[->] (21) -- (4);
\draw[->] (3.8,-0.4) -- (2.45,-1.15);
\draw[->] (31) -- (4);
\end{tikzpicture}
    \caption{An RDCEG graph $\mathcal{G}^{W-}$ for a bombing plot.}
    \label{fig:fig3}
\end{figure}
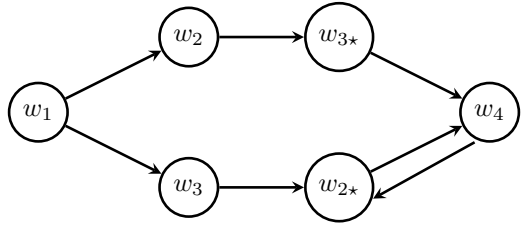

A possible RDCEG of the bombing plot is given in Figure \ref{fig:fig3}. Notice here this graph
embeds and depicts substantive elicited information about how it is believed
this particular plot might unfold. More explicitly it represents that - within the expected time scale of the plot - once trained ($w_2$) the agent does not forget his bomb making training ($w_{1}$), or the agent does not need to identify a target or
reconnoitre it again ($w_{3}$) once mobilised ($w_{4}$). Therefore there are no paths back into these states once they have been achieved. Some of this
information is almost a logical consequence of the meaning of the states -
others represent the expert opinion about this particular unfolding plot. On
the other hand, note that bomb making material can obviously be discarded once
acquired - returning the agent to a previous state and necessitating their
re-acquisition. This explains the transitions into and out of the state
representing the need for acquiring bomb materials. Note that it is plausible to assume here that both $A$ and $D$ are aware of this structural information.

\subsubsection{Transition Probabilities and Holding Times}
\label{ss:ht}
Each active state of this RDCEG is allocated a holding time distribution.
Although not strictly necessary, for notational simplicity, here we make a
fifth assumption that the time interval of the series is small enough that
the agent of the plot can make at most one transition into a successive
phase during a single time interval. We assume $D$ specifies that such an
event will happen with probability
\begin{equation}
q_{tj}, 0\leq q_{tj}\leq 1, \;j=1,2,\ldots ,m,\; t=1,2,\ldots ,T. 
\end{equation}
Independently of moving and whilst
temporarily staying in phase $w_{j}$, $D$ assumes that with elicited probability 
\begin{equation}
q_{tj}^{\prime }, \;0\leq q_{tj}^{\prime }\leq 1,\; j=1,2,\ldots ,m, \;t=1,2,\ldots ,T
\end{equation}
the agent may decide or be forced
to abandon the plot entirely. 
In the context of a
plot we can think of this latter event as the agent transitioning into the
inactive state $w_{0}.$ In the running example these probabilities will rely
on expert judgements about the nature of the activities, the past behaviour
of similar agents and the particular suspect being observed: all specific to
the particular plot under investigation.

Conditional on the agent transitioning to a subsequent active state, the
semantics of a RDCEG has a graph $\mathcal{G}^{W}$ where a directed edge
connects an active state $w_{i}$ to another active state $w_{j}$ - written $%
j\in E_{i}$ - if and only it is possible to transition from phase $w_{i}$ to 
$w_{j}$ within a single time step, $1\leq i,j\leq m$. It is this graph we
will concatenate to form the composite causal graph $\mathcal{G}$. We denote
the floret vector of probabilities of an agent moving to an adjacent active
phase, given they move and do not abort the mission by 
\begin{equation*}
\boldsymbol{p}_{ti}^{W}=\left( p_{tij}^{W}:w_{j}\text{ is connected from }w_{i}\text{ by an edge in }\mathcal{G}^{W}\right)
\end{equation*}%
Two vertices $w_{i}$ and $w_{j}$ and their associated edges are coloured the
same in $\mathcal{G}^{W}$ if their emanating edge probabilities are
isomorphic - i.e. $\boldsymbol{p}_{ti}^{W}=\boldsymbol{p}_{tj}^{W}$. When
this happens $w_{i}$ and $w_{j}$ are said to be in the same \emph{stage} $%
w_{k}^{\prime }$, $k=1,2,\ldots ,m^{\prime }.$ Notice we need only keep the
probabilities associated with each stage $k$, $k=1,2,\ldots ,K$. As with
the other probabilities these will again be informed by past data, the MO of the typical agent of $A$ and information about the observed suspect.

Although, since the RDCEG is generally formulated as a continuous time process,
within this application $D$'s inferences will only concern the phases in
which the agent resides in each of the designated time points. The
assumptions above then enable us to represent this by concatenating the
graph of the RDCEG with $\mathcal{G}^{B}$ and $\mathcal{G}^{S}$. Using the
notation and assumptions above, the probability mass function $p_{t_{0}}^{W}(%
\boldsymbol{w}_{t_{0}}^{T})$ is Markov and can be factorised as

\begin{equation}
p_{t_{0}}^{W}(\boldsymbol{w}_{t_{0}}^{T})=p_{t_{0}}^{W}(w_{jt})\prod%
\limits_{t=t_{0}+1}^{T}p_{t}^{W}(w_{jt}|w_{it-1})  \label{W formula}
\end{equation}%
where for $t=t_{0}+1,t_{0}+2,\ldots ,T$ 
\begin{equation*}
p_{t}^{W}(w_{jt}|w_{0t-1})=\left\{ 
\begin{array}{cc}
1 & w_{jt}=w_{0} \\ 
0 & w_{jt}\neq w_{0}%
\end{array}%
\right.
\end{equation*}%
and for $i=1,2,\ldots ,m$

\begin{equation}
p_{t}^{W}(w_{jt}|w_{it-1})=\left\{ 
\begin{array}{cc}
q_{ti}^{\prime } & j=0 \\ 
(1-q_{ti}^{\prime })(1-q_{ti}) & 1\leq i=j\leq m \\ 
(1-q_{ti}^{\prime }) q_{ti} \;p_{tij}^{W} & 1\leq i\neq j\leq m,j\in E_{i}
\\ 
0 & 1\leq i\neq j\leq m,j\notin E_{i}%
\end{array}%
\right.
\end{equation}%
Therefore, the set of factors associated with this deepest level of the
hierarchical model is 
\begin{equation*}
\mathcal{P}_{\mathcal{G}}^{W}=\left\{ p_{t_{0}}^{W},q_{ti}^{\prime
},q_{ti},p_{tkj}^{W}:1\leq i\leq m,j\in E_{i},\text{ }k=0,1,2,\ldots
,K, \; t=t_{0},t_{0}+1\ldots ,T\right\}
\end{equation*}

\subsection{The Structural Graph of an Idle Plot}

We can now demonstrate that the class of plot models is graphical in the
sense of the definition above. The coloured graph $\mathcal{G}$ framing
our causal algebra is a hybrid of a RDCEG graph, a bipartite graph, and the
graph of a 2TDBN and is given by 
\begin{equation*}
\mathcal{G}\triangleq \mathcal{G}^{W}\cup \mathcal{G}^{B}\cup \mathcal{G}^{S}
\end{equation*}%
Here the union operation denotes the union of the vertices and edge sets of
these graphs. Any colouring on vertices is inherited from $\mathcal{G}^{W}.$
The semantics $\mathcal{S}_{\mathbb{G}}$ of the composite graph of a plot
model is inherited from the established semantics of the corresponding three
component graphs described above. Note that the information embedded in each
component graph is structural and can be explained in natural language. The
hybrid coloured graph $\mathcal{G}$ embodies the three critical components
of a plot model; this can be used to unambiguously frame a family of
probability models consistent with the structural information conveyed
through its topology and colouring. Such intrinsic structural information
makes the hybrid graph ideal for describing the causal extension we develop below. 

Once the appropriate graph $\mathcal{G}$ has been elicited from $D$, they need to provide the factors in $\mathcal{P}_{\mathcal{G}}$ determined by
this graph - using the formulae above - for the joint distribution of $%
\boldsymbol{\xi }_{1}$ - described by the joint density/mass function $p_{1}(%
\boldsymbol{\xi }_{1})$ - of the idle system to be fully specified.

These required factors are 
\begin{equation*}
\mathcal{P}_{\mathcal{G}}\mathcal{\triangleq P}_{\mathcal{G}}^{W}\cup 
\mathcal{P}_{\mathcal{G}}^{B}\cup \mathcal{P}_{\mathcal{G}}^{S}
\end{equation*}%
where $\mathcal{P}_{\mathcal{G}}^{W},\mathcal{P}_{\mathcal{G}}^{B},\mathcal{P%
}_{\mathcal{G}}^{S\text{ }}$ are as defined above. Our defined concatenation
of the graphical model then implies that the joint density $p_{t_{0}}(%
\boldsymbol{\xi })$ of all phases, their tasks and intensities between times 
$t_{0}$ and $T$ can be written as a function of factors in $\mathcal{P}_{%
\mathcal{G}}$. Thus the density of any idle plot model can be given by the
product formulae 
\begin{eqnarray}
p_{t_{0}}(\boldsymbol{\xi }_{t_{0}}) &=&p(\boldsymbol{z}_{t_{0}}^{T},%
\boldsymbol{\theta }_{t_{0}}^{T},\boldsymbol{w}_{t_{0}}^{T})=f_{\mathcal{G}}(%
\boldsymbol{p}^{S},\boldsymbol{p}^{B},\boldsymbol{p}^{W})\triangleq f_{S}(%
\boldsymbol{p}^{S})f_{B}(\boldsymbol{p}^{B})f_{W}(\boldsymbol{p}^{W})  \notag
\\
&=&p_{t}^{S}(\boldsymbol{z}_{t}|\boldsymbol{\theta }_{t},\boldsymbol{z}%
^{t-1},\boldsymbol{\theta }^{t-1})p_{t}^{B}(\boldsymbol{\theta }_{t}|%
\boldsymbol{z}^{t-1},\boldsymbol{\theta }^{t-1},w_{t_{0}})f_{W}(\boldsymbol{p%
}^{W})  \label{function formula}
\end{eqnarray}%
where $f_{S}(\boldsymbol{p}^{S}),f_{B}(\boldsymbol{p}^{B}),f_{W}(\boldsymbol{%
p}^{W}),$ are themselves functions of only factors $\boldsymbol{p}^{S}\in 
\mathcal{P}_{\mathcal{G}}^{S}$, $\boldsymbol{p}^{B}\in \mathcal{P}_{\mathcal{%
G}}^{B}$ and $\boldsymbol{p}^{W}\in \mathcal{P}_{\mathcal{G}}^{W}$
respectively. Here - for clarity - we have suppressed the arguments of these
functions. We note from Equation (\pageref{observed densities}) that $f_{S}$ is a function of products in $\mathcal{P}_{\mathcal{G}}^{S}$ and from Equation (\ref{wi B factors}) that $f_{B}(\boldsymbol{p}^{B})$ is a
function of products of factors in $\mathcal{P}_{\mathcal{G}}^{B}$. Finally, $%
f_{W}(\boldsymbol{p}^{W})$ is a function of the products given in Equation (\ref{W
formula}).

Since the structural form of the graph $\mathcal{G}$ of a plot describes
how $D$ believes the fundamental process advances, as with more conventional
causal graphs, $D$ is able to use it to represent the impacts of
interventions $d\in \mathbb{D}$ that do not invoke an intelligent reaction
from $A$ which we demonstrate and illustrate below.

\subsection{An Illustrative Example}
\label{subsec:example}
As part of our work, we have developed a bespoke Python library to make inference for problems that are well-modelled by the class of plot models.  Using this software, it is possible to implement the hypothetical plot introduced in Example 1.

Applying our software for this example is a question of specifying the various factors, as outlined in Section \ref{subsec:plot_specification}.
In the absence of domain experts from whom we can elicit relevant information to populate the required conditional probability distributions, we construct our model based on approximate estimates of such factors. While such an approach would not be appropriate for operational use, it is sufficient for a proof-of-concept illustration. 

Given the model specification, our software allows us to simulate a synthetic time series of observations that is representative of an adversary undertaking such a plot. Treating this data as observations from a real plot, we can apply a filtering propagation algorithm, allowing the observations to update $D$'s inferred estimates of the probability that the adversary is in each phase of the plot at each time point. 

\begin{figure}
\includegraphics[width=\textwidth]{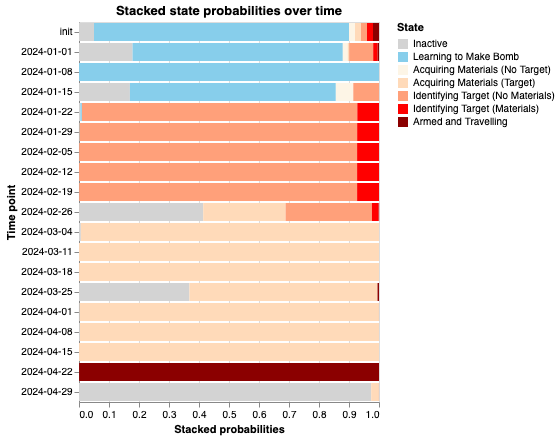}
\caption{A stacked bar-chart of the evolution of probability estimates of the plot phase of the adversary $A$, based on filtering observed data at each time point. \label{fig:filter_visual}
}
\centering
\end{figure}

Figure \ref{fig:filter_visual} shows a stacked bar-chart, with time on the Y-axis (increasing moving down the axis). For each time point, the horizontal bar displays $D$'s estimates of the probability that $A$ is in each phase of the plot, updated to account for the observed data until that time point. As can be seen, while there is some uncertainty for a few time points, for this simplified and synthetic example, the observed data and specified model dynamics typically combine to allow the defender $D$ to estimate a single plot phase as the most probable at each time. For example, for this particular simulation, it is clear that the adversary sought to identify the target (weeks 2023-01-22 to 2024-02-19) before acquiring the necessary material to undertake the attack (weeks 2024-03-26 to 2024-04-15). There is more uncertainty about the activity that the attacker was undertaking during the week commencing 2024-02-26.

\section{Causal Algebras for Plots}

\subsection{A Typology of Defensive Interventions}

For causal analyses it is often useful to represent the impact of considered
interventions in terms of composites of simpler interventions whose causal
algebras are most transparent. So, for example in the case of BNs, it has
proved useful to first study the effects of ``do" operators and only then express the effects of more complex interventions as functions of these.
Here, because of the hybrid nature of the underlying graph of plots, unlike
for BNs, it is necessary to consider a variety of simpler interventions that
correspond to the three different levels
of its hierarchy. We then add a fourth which concerns $A$ switching away
from their current plot into another activity:

\begin{enumerate}
\item\textbf{Clarifying Interventions} \\ 
Here the defender $D$ can intervene at the 
\textbf{surface level} $d\in \mathbb{D}_{S}$ of a plot model. Such
interventions typically modify factors in $\mathcal{P}_{d\mathcal{G}}^{S}$
and affect the \emph{quality} of information police can obtain about a
suspected plot. Examples within the bombing example above include obtaining
permission to collect sensitive data from the suspect's associates,
infiltrating his friendship grouping or installing covert cameras to better
observe the tasks being undertaken. Each such $d\in \mathbb{D}_{S}$ is
designed to improve $D$'s information about the progress of a plot. Note
that governments can also intervene to inhibit information police can
collect, for example by passing new laws to restrict access to certain
personal data.

\item\textbf{Blocking Interventions}\\ 
Here the defender $D$ intervenes at the 
\textbf{task level} $d\in \mathbb{D}_{B}$ of a plot model. Such
interventions typically modify factors in $\mathcal{P}_{d\mathcal{G}}^{B}.$
Examples include electronically or physically disrupting the suspect's
communications with associates of $A$, making it more difficult to perform
the tasks needed to move to the next phase of the plot using information
they currently have about $A$'s MO. These interventions typically inhibit and
discourage $A$ from achieving their objective. So, in the example above, $D$
could try to inhibit the agent's bomb making training by preventing the
suspect from travelling to another country legitimately - for example by
confiscating his passport. We note that an unintelligent reaction to this
would be for the agent to give up obtaining training in their planned way -
here by travelling abroad. More usually an agent is likely to
react intelligently and find some other way to acquire the necessary skills.

\item\textbf{Direct Interventions} \\ Here the defender $D$ intervenes at the \textbf{%
phase level} $d\in \mathbb{D}_{W}$ and attempts to frustrate a suspect at a
given phase of attack, thus preventing the plot from proceeding further.
Such interventions typically modify factors in $\mathcal{P}_{d\mathcal{G}%
}^{W}.$ In the bomb attack example, one intervention might be to precipitate
a police raid on a property where bomb making is believed to be happening or
bombs have already been made. Another might be to kill a suspect who is armed with a
bomb. Another might be for $D$ to inform $A$'s agent that they are aware the
plot is underway and that, if it were to proceed, then the agent would be
arrested.

\item\textbf{Disabling Interventions}\\ Here the defender $D$ can intervene at the 
\textbf{plot level} $d\in \mathbb{D}_{P}$ by making the currently enacted
plot so difficult to perpetrate it forces $A$ to switch to return to $w_{0}.$
For example, in the terrorism example the suspect is told that authorities
realise that he is about to perpetrate a bombing plot and if he does not abort
immediately then he will be arrested. This type of intervention is designed to
prevent the type of plot being perpetrated and force the suspect to either
give up or switch to perpetrating a different attack (such as a knife
attack).
\end{enumerate}

Of course, for an algebra to be compelling it needs to be customised to its
context. For the remainder of this section, we will illustrate the
implementation of the construction of a customised causal algebra using our
running example of a bombing plot. Here we simply consider just one typical
intervention chosen from each of $\mathbb{D}_{S}$, $\mathbb{D}_{B}$, $\mathbb{D}_{W}$
and $\mathbb{D}_{P}$.

\subsection{Causation Not Invoking an Intelligent Reaction}

We next demonstrate how $D$ can construct a compelling causal algebra around
a selected $\mathcal{G\in }$ $\mathbb{G}$ that predicts the impacts of
typical interventions $d\in \mathbb{D}$ that $D$ might contemplate enacting to
frustrate a plot they believe is underway. We begin by discussing the
simplest scenario where $D$ believes $A$ is unable or unwilling to react to
their intervention $d\in \mathbb{D}$. Then in later sections we will use ARA
technologies adapted to plot models to extend these scenarios to ones where $%
A$ can intelligently react to such simple interventions. Such scenarios are
more complex since they often demand that $\mathcal{G\in }$ $\mathbb{G}$ is
embellished so that it can describe features associated with $A$'s intent,
knowledge and capability, which will determine how they might act to mitigate
an intervention. We show below that these sorts of extensions for causal graphs - first
discussed in \cite{Ramiahetal23} - are quite straightforward to embed for
plot models.

First we demonstrate how $D$ can use the graph $\mathcal{G\in }$ $\mathbb{G}$
of the idle system to construct the density of the effects of a particular
intervention on an unreactive adversary $A$. In analogy with a BN, this
simply involves retaining the structural form $\mathcal{G\in }$ $\mathbb{G}$
of the plot model, substituting a few new factors from $\mathcal{P}_{%
\mathcal{G}}$ specifically associated with the particular intervention $d\in 
\mathbb{D}$ and retaining the other factors in $\mathcal{P}_{\mathcal{G}}$.
We denote this new set of factors by $\mathcal{P}_{d\mathcal{G}}$. We then
construct the intervened density $p_{dt_{0}}(\boldsymbol{\xi }_{t_{0}})$
needed to calculate our expected utility scores in an SEU analysis using $%
\mathcal{G}$ and its formula in Equation (\ref{function formula}) with the new set of
factors associated with this intervention. For unrevealed interventions the
appropriate algebras appear to be ones which are analogous, but distinct from
those already developed for various dynamic versions of BNs and CEGs, and now applied to this hybrid graph.

\subsection{Unintelligent Interventions in a Bombing Plot}

We next illustrate the causal algebras associated with just four different
types of interventions that might be made by $D$ during the bombing attack
discussed above.

\begin{enumerate}
\item $d_{s}\in \mathbb{D}_{S}$. An agent of $D$ has just managed to
befriend $A$'s agent who is now able to inform $D$ with more precision about
certain tasks $A$'s agent is currently engaging in. $D$ receives higher quality information the longer the friendship lasts. We can introduce
information expressed by adding into the graph a new component to
the intensity vector $\boldsymbol{Z}_{i}$ of tasks $\theta _{i}$ for a
selection of the tasks $i\in I_{d_{s}}$. Under the notation above this
replaces the current sample densities 
\begin{equation*}
p_{ti}^{S}(\boldsymbol{z}_{ti}|\theta _{ti},\boldsymbol{z}_{tQ^{\prime
\prime }(i)}):i\in I_{d_{s}},t=t_{0},t_{0}+1,\ldots ,t_{1}(d_{s})
\end{equation*}%
with new more informative sample densities 
\begin{equation}
p_{d_{s}ti}^{S}(\boldsymbol{z}_{ti}|\theta _{ti},\boldsymbol{z}_{tQ^{\prime
\prime }(i)}):i\in I_{d_{s}},t=t_{0},t_{0}+1,\ldots ,t_{1}(d_{s})
\label{subsurface}
\end{equation}%
where $t_{1}(d_{s})\leq T$ represents the time at which the friendship ends.
Depending on the nature of the friendship in our running example, the tasks the agent could inform $D$ about are, say, in the
index set $I_{d_{s}}\triangleq \left\{ 1,3,5,7\right\} $. Thus in such a
setting our causal map $\mathcal{P}^{S}\rightarrow \mathcal{P}_{d_{s}}^{S}$
is defined by Equation (\ref{subsurface}) whilst $\mathcal{P}_{d_{s}}^{B}=\mathcal{P}%
^{B}$ and $\mathcal{P}_{d_{s}}^{W}=\mathcal{P}^{W}$ remain invariant under
this map.

\item $d_{b}\in \mathbb{D}_{B}$. The defender confiscates the agent's
passport purportedly because that agent is a suspect in a completely
unrelated crime. Suppose the agent is unable or unwilling to act on their
own initiative and deviate from an original plan because such an
intervention would prevent the agent from travelling legitimately. So this
obedient or un-resourceful agent who had been ordered to go abroad for
training would then be forced to abort the plot if they had not yet received
training at time $t_{0}$. In our notation from Section \ref{ss:ht}, this would set $q_{d_{b}1t}^{\prime }=1$, 
$t=t_{0},t_{0}+1,\ldots, T$ 
keeping all other probabilities invariant. For the reactions of an intelligent agent, see Section \ref{sec:5}.

\item $d_{w}\in \mathbb{D}_{W}.$ The defender $D$ connects with $A$'s agent
and tells them they are aware of the plot. By definition this is a case
where $A$'s intelligence is triggered and so will be discussed in Section \ref{sec:5}.

\item $d_{p}\in \mathbb{D}_{P}.$ The defender prepares a submission to
extradite the suspect to another country. Although the agent might re-enter
the country illegitimately later, if the submission is successfully applied
then the current plot will be foiled.
\end{enumerate}

Note that although the causal algebras associated with these different
interventions are a little more complicated than, for example, those
typically studied for BNs or CEGs and need more contextual inputs and
domain judgements before they are specified, they are nevertheless
transparent and relatively straightforward to implement.

\section{Intelligent Causal Algebras for Plots}\label{sec:5}

\subsection{Introduction}

In the last section we demonstrated that until an adversary has learnt that an
intervention has happened, the implications of the causal algebras were
broadly analogous to the algebras of the do calculus of more standard
models. However, we also illustrated that, because of the sequential nature
of a plot, for most interventions that $D$ might contemplate, any causal
algebra would need to model how $A$ might react once they heard of the
intervention. For example, even for a simple intervention like removing an
agent's passport, if this leads to $A$ suspecting this has happened because $%
D$ had discovered their plot then $D$ should surely adjust - through their
selected causal algebra - their beliefs about how $A$ will react to this
discovery and so their beliefs about how that plot might unfold. Even more
obviously, if $D$ raids a premises before incriminating evidence has accrued
- in our case before the bomb had actually been manufactured so that no
arrest could be made - then it will be obvious to $A$'s agent that $D$ has
discovered the plot and is able to react to this.

In such circumstances $D$ will then need to model \emph{why} and \emph{how}
the rational adversary acts, not simply \emph{what }they might do - $D$ would need to consider all of these factors when modelling the idle system. So $D$ may need
to build into the topology of their causal graph $\mathcal{G}$ of the idle
process extra features that explain the ``why" and ``how",  which reflect $A$'s intent and capabilities to react. Therefore $D$ will often need to
embellish the graph $\mathcal{G}$ into a new graph $\mathcal{G}^{+}$ to
embed these new features - latent in the idle model, but active for some $%
d\in \mathbb{D}$: see \cite{Banksbook,Banks22,Rios09,Rios16,Rios23}. How this might apply for a class of models,
simpler than the plot models, has already been illustrated in \cite{Ramiahetal23}.

We implement and then illustrate these ideas as they apply to plots below.
Typically, there are two sorts of discovery within a plot that $A$ can make through
the intervention $d\in \mathbb{D}$ that $D$ might enact:

\begin{enumerate}
\item \emph{Local discovery}: $A$ becomes aware that $D$ has intervened
using methods $\mathbb{D}_{S},\mathbb{D}_{B}$, $\mathbb{D}_{W}$
and $\mathbb{D}_{P}$.

\item \emph{Plot discovery}: $A$ becomes aware that $D$ has discovered their
agent's plot is underway. This second case is often automatic - i.e. will
happen with probability $1$ if $D$ chooses to intervene using an intervention
either in $\mathbb{D}_{W}$ or $\mathbb{D}_{P}$
\end{enumerate}

Note that the second is what is sometimes referred to as a second level
inference because it engages a deeper level of an infinite regress in any
game theoretic formulation of this process \cite{Allpranisgame}.
Typically, in practice, any intervention by $D$ has a non-zero probability
that it will betray to $A$, $D$'s awareness of $A$'s plot. This means that
sometimes different features need to be added to describe how $A$ might
react to counter either of these two types of discovery.

Once in place, if embellished wisely, $\mathcal{G}^{+}$ can then be used as
the common causal framework that is able to not only describe the idle
system (now in more detail) but also the impact of the decisions $d\in 
\mathbb{D}$ that the defender $D$ might contemplate making. In particular, $D$
will only need to find ways of systematically specifying the new factors $%
\mathcal{P}_{d\mathcal{G}^{+}}$ associated with each $d\in \mathbb{D}$. They
can then use the formula for the plot associated with $\mathcal{G}^{+}$ to
calculate $D$'s SEU scores as we demonstrate below.

The next question is how $D$ might \emph{formally} construct $\mathcal{G}^{+}
$ from $\mathcal{G}$ by modelling $A$'s likely reaction to their intervention?
In \cite{Ramiahetal23} we argued that in our Bayesian setting, to develop
such causal algebras, we should follow the now established ARA protocols. In the form of ARA we adopt here, the defender $D$
assumes that $A$ is rational in the sense of being an SEU maximiser too. Moreover, the defender $D$
assumes that $A$ reacts intelligently to the two types of discovery outlined
above, consistently with their intent and capabilities. By making this
assumption $D$ is then able to put themselves in $A$'s shoes and model
predictively how $A$ is likely to react. Our new contribution here is that -
although what $A$ believes might be uncertain to $D$ - in a setting where $D$
believes that their chosen graphical model $\mathcal{G}^{+}$ is \emph{causal 
}- the intelligent adversary $A$ will also share the structural insights
embedded in $\mathcal{G}$. This assumption can then be embedded in any ARA\
analysis of a plot.

We note that it has long been appreciated that any system labelled as
"causal" must embed the idea that the mechanisms it describes is appreciated
as plausible and ideally shared by other intelligent people, see e.g. \cite{BradfordHill,WardBradHillCause}. It would follow that, in the
context of graphical models, if $D$ assumes $\mathcal{G}^{+}$ is truly
causal then it is at least plausible for them to believe that $\mathcal{G}%
^{+}$ also expresses $A$'s beliefs. However, once $D$ makes this causal leap of faith, $D$'s task of double guessing $A$'s beliefs and thus their rational
reactions within an ARA analysis is greatly simplified. Then $D$ only needs to specify what they believe $A$ will use for their factors on this graph. The
only difference here is that - unlike $D$ - $A$ is likely to know how their agent
will react and so the corresponding conditional probability factors will then be
deterministic rather than stochastic.

Henceforth in this section we use the following notation:

\begin{enumerate}
\item Random vectors, denoted by $\left\{ \boldsymbol{K}_{d},d\in \mathbb{D}%
\right\} $, that are uncertain to $D$ but known to $A$, measure when and to what
extent $D$ believes $A$ can discover each contemplated $d\in \mathbb{D}$.
For plots this may include the two types of discovery listed above.

\item A random variable $U_{A}(\boldsymbol{y})$ over a vector of attributes $%
\boldsymbol{y}$ - uncertain to $D$ but known to $A$ - is the SEU
maximisers of $A$'s utility function expressing $A$'s intent that applies to
the current plot. Typically $D$ will be informed about the nature of $U_{A}(%
\boldsymbol{y})$ from what $A$ pronounces, how they have behaved in the past
and from $D$'s intelligence, but usually not simply from data about the idle
system. In our running example, for simplicity we assume that $D$ believes
this is a three state function 
\begin{equation}
\left\{ 
\begin{array}{cc}
1 & \text{if the plot is successful} \\ 
u_{A} & \text{if the agent aborts the plot but escapes arrest or death} \\ 
0 & \text{if the agent is disabled before the plot is completed}%
\end{array}%
\right.   \label{A's utlility}
\end{equation}%
where $0\leq u_{A}\leq 1$.

\item Random vectors - here denoted by $\left\{ \boldsymbol{C}_{d}:d\in 
\mathbb{D}\right\} $ - again known to $A$ but not $D$. This vector can be
used by $D$ to express their beliefs about how $A$ might be able to react
having discovered $D$'s possible interventions in order to mitigate their
impact. Such capabilities typically need to express the ability and freedom $%
A$'s agent might have to react as well as their physical and skill based
constraints. For example, if an agent is programmed simply to obey orders to
act then they must receive and understand orders before they can enact them. On
the other hand, if they are given freedom to act unilaterally on behalf of $A$
then their reactions can be expected to be more agile. All such
considerations need to be folded into $D$'s subjective distribution of $%
\left\{ \boldsymbol{C}_{d}:d\in \mathbb{D}\right\} $. Within a plot such
vectors typically relate to alternative tasks that $A$ might apply to
circumvent $D$ when they try to block the progress of the plot at a given
phase.
\end{enumerate}

These three newly introduced components and the training $D$ believes the
agent has received - which will define their MO and the space of reactions
they choose from - will then frame the agent's potential chosen reactions $%
\boldsymbol{R}_{d}^{\ast }$ to each intervention $d\in \mathbb{D}$.

Our suggested protocol for constructing an intelligent causal algebra - as
this might apply to a plot - then proceeds as follows:

\begin{enumerate}
\item $D$ examines and, if necessary, embellishes the causal structure of the
plot to ensure that the chosen graph $\mathcal{G}$ is sufficiently
expressive to embed the salient information $D$ has about $\left\{ \left( 
\boldsymbol{K}_{d},U_{A},\boldsymbol{C}_{d},\boldsymbol{R}_{d}^{\ast
}\right) :d\in \mathbb{D}\right\} $ described above.

\item $D$ embeds the information about the factors $\mathcal{P}_{d\mathcal{G}%
}\triangleq \left\{ p_{d1\mathcal{G}}(\boldsymbol{\xi }),p_{d2\mathcal{G}}(%
\boldsymbol{\xi }),\ldots ,p_{dm\mathcal{G}}(\boldsymbol{\xi }):d\in \mathbb{%
D}\right\} $ they will be using. If $\mathcal{G}$ has been embellished then
we first check that the new factors give predictive distributions over
effects that are consistent with those of the original idle model. $D$ then
chooses what they believe factors associated with the triple $\left\{ \left( 
\boldsymbol{K}_{d},U_{A},\boldsymbol{C}_{d}\right) :d\in \mathbb{D}\right\} $
described above will be.

\item Since $D$ believes $\mathcal{G}$ is causal they believe that $A$ will
reason consistently with $\mathcal{G}$. So in particular $A$ will choose a
set of factors 
\begin{equation*}
\mathcal{P}_{d\mathcal{G}}^{A}\triangleq \left\{ p_{d1\mathcal{G}}^{A}(%
\boldsymbol{\xi }),p_{d2\mathcal{G}}^{A}(\boldsymbol{\xi }),\ldots ,p_{dm%
\mathcal{G}}^{A}(\boldsymbol{\xi }):d\in \mathbb{D}\right\}
\end{equation*}%
to enable them to calculate their density $p_{\boldsymbol{r}(d)}^{A}(%
\boldsymbol{y}|\boldsymbol{K}_{d},U_{A},\boldsymbol{C}_{d})$ over their
attributes. Of course, by definition $\left\{ (\boldsymbol{K}_{d},U_{A},%
\boldsymbol{C}_{d}):d\in \mathbb{D}\right\} $ will be known to $A$, but usually be latent to $D$. We argue that in many settings for unintelligent
conditioning, $D$ can plausibly assume that many of the factors in $\mathcal{P%
}_{d\mathcal{G}}^{A}$ will be identical to the corresponding factors in $%
\mathcal{P}_{d\mathcal{G}}$.

\item To predict factors describing $(\boldsymbol{R}_{d}^{\ast }|\boldsymbol{%
K}_{d},U_{A},\boldsymbol{C}_{d})$, for each intervention $d\in \mathbb{D}$, $%
D$ then uses an ARA analysis \cite{JoshiRios,RiosInsua}.
Thus because $D$ believes $A$ to be an SEU decision maker they assume $A$
will maximise%
\begin{equation}
\overline{U}_{A}\left( \boldsymbol{r}_{d}\right) \triangleq \int U_{A}\left( 
\boldsymbol{y}\right) p_{\boldsymbol{r}_{d}}^{A}(\boldsymbol{y}|\boldsymbol{K%
}_{d},U_{A},\boldsymbol{C}_{d})d\boldsymbol{y}  \label{A's score}
\end{equation}%
where $\left\{ p_{\boldsymbol{r}_{d}}^{A}(\boldsymbol{y}|\boldsymbol{K}%
_{d},U_{A},\boldsymbol{C}_{d}):d\in \mathbb{D}\right\} $ are $A$'s
conditional densities over their attributes for different decisions. So $A$
chooses a reaction 
\begin{equation*}
\boldsymbol{r}^{\ast }(d|\boldsymbol{K}_{d},U_{A},\boldsymbol{C}%
_{d})\triangleq \arg \max_{r_{d}\in \mathbb{R(}d)}\overline{U}^{\ast }\left( 
\boldsymbol{r}_{d}|\boldsymbol{K}_{d},U_{A},\boldsymbol{C}_{d}\right) 
\end{equation*}%
Note here the flexibility of this ARA approach. Thus if $A$ believes that $D$
might not be fully rational (for example, have a "trembling hand" \cite{Allpranisgame}) then their distribution of $\boldsymbol{r}_{d}^{\ast }|p_{%
\boldsymbol{r}_{d}}^{A}(\boldsymbol{y}|\boldsymbol{K}_{d},U_{A},\boldsymbol{C%
}_{d})$ may not be deterministic. The defender can simply substitute a
distribution reflecting what they believe $A$'s deviation from rationality
might be.

\item We note that $D$'s scores $\left\{ \overline{U}_{D}\left( d\right)
:d\in \mathbb{D}\right\} $ will depend only on $D$'s margin over what reactions 
$\left\{ \boldsymbol{r}_{d}:d\in \mathbb{D}\right\} $ $A$ might choose and not
the reasons for this choice as measured through $\left( \boldsymbol{K}%
_{d},U_{A},\boldsymbol{C}_{d}\right) $: see e.g. \cite{Ramiahetal23}.
\end{enumerate}

The critical issue in using an ARA methodology in general is how $D$
constructs $$\left\{ p_{\boldsymbol{r}(d)}^{A}(\boldsymbol{y}|\boldsymbol{K}%
_{d},U_{A},\boldsymbol{C}_{A}):d\in \mathbb{D}\right\} $$ so as to evaluate Equation (\ref{A's score}). Several methods of doing this have been proposed \cite{JoshiRios,RiosInsua}, each appropriate for different kinds of
inferences. However, the extended definition of causation used in \cite%
{Ramiahetal23} greatly simplifies this otherwise much more delicate
modelling process. This is because the shared causal graph enables $D$ to
break this inferential process down into manageable substeps. We illustrate
this process below for plots.

\subsection{A Taxonomy of Interventions in Intelligent Plots}

\subsubsection{On Modelling an Intelligent Adversary Generically}

Since plot models are, by their very nature, highly descriptive of what
and why $A$ might be acting the way they are, unlike some other classes of
graphical models, since the difference between $\mathcal{G}^{+}$ and $\mathcal{G}$
is not too large. For example, $A$'s utility when engaging in a plot is
already explicitly modelled through $D$'s use of a plot model. So this
hypothesis is already embedded when $D$ believes that $A$'s utility takes
the same form as Equation (\ref{A's utlility}). Furthermore, for $D$ to model $A$'s capability
to intelligently respond, $D$ only needs to ensure that their description of the
task vector is rich enough to not only include $A$'s preferred way of
progressing their plot - expressing their current MO and exhibited through
their acts within the idle system - it now also needs to include descriptions
of alternative ways $A$ might progress their intent if $D$ chooses to
intervene in a way that inhibits their preferred MO.

We first ignore the impact of plot betrayal and adapt $\mathcal{G}$ to a
graph $\mathcal{G}^{\prime }$ that captures ways in which $D$ believes $A$
is capable of modifying their current MO to circumvent the impacts of
different interventions $d\in \mathbb{D}$. A second map then describes how $%
D $ might further modify their causal description now embedded in $\mathcal{G%
}^{\prime }$ to produce a causal graph $\mathcal{G}^{+}$ that can capture this potential ``second level" \cite{Allpranisgame} impact of a given $d\in \mathbb{D}$ as well. We illustrate below that it is not unusual - if
the task sets are defined richly enough initially - for $\mathcal{G}^{\prime
}=\mathcal{G}$.

An intelligent $A$ might react to different interventions by $D$ - driven by
their intent described by Equation (\ref{A's utlility}) in a number of ways:

\begin{enumerate}
\item If $A$ discovers interventions by $D$ at the \textbf{surface level} $%
d\in \mathbb{D}_{S}$ of a plot model then $A$ can, at a local level, adapt their
behaviour by either \emph{disguising} their acts or actively \emph{trying to
deceive} $D$ and misinterpret what $D$ sees to advantage $A$.

\item If $A$ becomes aware of an intervention by $D$ at the \textbf{task
level} $d\in \mathbb{D}_{B}$ of a plot model then a rational $A$ will react
by harnessing all their capabilities to \emph{modify} the way they plan to
move through a phase by choosing alternative tasks to achieve their ends
within a given phase. Alternatively, they might continue performing the same
tasks, but in a way that it is less easy for $D$ to see.

\item When $A$ learns that $D$ plans to intervene at the \textbf{phase level}
$d\in \mathbb{D}_{W}$ they usually implicitly learn that their plot had been
discovered. They can react to avoid its worse consequences by \emph{resetting%
}, aborting the phase completely and return themselves to an earlier phase.
Alternatively, they could disguise the tasks they are performing to complete
the phase, modify the types of tasks they are undertaking or simply
abort their plot.

\item If the defender $D$ successfully intervenes at the \textbf{plot level} 
$d\in \mathbb{D}_{P}$ then $A$'s agent is forced, in the medium term, to
become inactive $w_{0}$ and abort the plot. The effect of this type of
intervention is simple and their effect needs only a trivial adaptation of
probabiiities in the idle process. 
\end{enumerate}

\subsubsection{Intelligent Local Reactions to Different Levels of Intervention}
\label{subsubsec:intelligent}

We can now examine how $D$ might reason the effects an intervention might
have if they assume $A$ can react intelligently. We begin by considering a
setting where $D$ believes that $A$ - although aware of the intervention -
will not suspect that the plot had been discovered:

\begin{enumerate}
\item $d_{s}\in \mathbb{D}_{S}$. $A$'s agent begins to distrust the
introduced befriender. They either decide to end the friendship or look to
actively deceive the befriender and feed them with disinformation. The
original graph is sufficient to provide a frame for these possibilities. The
defender $D$ simply needs to modify some of the intensity densities. In the
first case the time of the discovery will be known by $D$, and $D$
simply reverts to the earlier sample densities in Equation (\ref{subsurface}). In the second case $D$ would
need to replace these densities by other densities reflecting their possible
purposeful contamination.

\item $d_{b}\in \mathbb{D}_{B}$. When the defender confiscates the agent's
passport purportedly because that agent is a suspect in a completely
unrelated crime. Suppose the agent is able to respond intelligently and
willing to adapt their plans. Suppose the underlying causal graph $\mathcal{G%
}^{\prime }$ has been embellished sufficiently to include all possible
alternative tasks $A$'s agent has available to him to try to circumnavigate
this type of intervention. Then, provided the agent fails to appreciate the
real reason for the confiscation, $D$ might want to make the following
adjustments to their model:

\begin{itemize}
\item $D$'s probabilities - factors in $\mathcal{P}^{W}$- that the agent
aborts might still increase. So $D$ resets $1\geq q_{d_{b}1t}^{\prime }\geq
q_{1t}^{\prime }$, $t=t_{0},t_{0}+1,\ldots, T$.

\item $D$'s probabilities - also factors in $\mathcal{P}^{W}$- that the agent will transition
through $w_{1}$ will be expected to increase - so that in our
notation $1\geq q_{d_{b}1t}\geq q_{1t}$, $t=t_{0},t_{0}+1,\ldots, T$.
\end{itemize}

The probabilities $D$ assigns to the agent's choice of tasks will obviously
also change after the passport has been confiscated. For example, suppose $D$
believes that $A$'s agent will be unable or unwilling to travel abroad and
return illegally. Then one default would be to apply Pearl's usual ``do"
algebra to the associated probabilities in $\mathcal{P}^{B}$. Then the
probabilities $A$ chooses will have the travel option $\theta _{3t}$, $t=t_{0}, t_{0}+1,\ldots,T$, given he does not abort the mission, will be $0$, whilst
under Pearl's default ``do" algebra applied to the 2TDBN on the complement of $\theta
_{3}$ of the intermediate layer of the plot, the probability $D$ chooses
option $\theta _{1}$ or $\theta _{2}$ would scale up. So then $D$ might
reset the factors 
\begin{eqnarray*}
p_{d_{b}t11}^{B}(\theta _{t1}|\boldsymbol{\theta }_{tQ(1),},\boldsymbol{%
\theta }_{(t-1)Q^{\prime }(1),}w_{1}) &\triangleq &Cp_{t11}^{B}(\theta _{t1}|%
\boldsymbol{\theta }_{tQ(1),},\boldsymbol{\theta }_{(t-1)Q^{\prime
}(1),}w_{1}) \\
p_{d_{b}t21}^{B}(\theta _{t2}|\boldsymbol{\theta }_{tQ(2),},\boldsymbol{%
\theta }_{(t-1)Q^{\prime }(2),}w_{1}) &\triangleq &Cp_{t21}^{B}(\theta _{t2}|%
\boldsymbol{\theta }_{tQ(2),},\boldsymbol{\theta }_{(t-1)Q^{\prime
}(2),}w_{1}) \\
p_{d_{b}t31}^{B}(\theta _{t3}|\boldsymbol{\theta }_{tQ(3),},\boldsymbol{%
\theta }_{(t-1)Q^{\prime }(3),}w_{1}) &\triangleq &\left\{ 
\begin{array}{cc}
1 & \textrm {if }\theta _{t3}=0 \\ 
0 & \textrm {if }\theta _{t3}=1%
\end{array}%
\right.
\end{eqnarray*}%
where 
\begin{equation*}
C\triangleq \left[ p_{t11}^{B}(\theta _{t1}|\boldsymbol{\theta }_{tQ(1),},%
\boldsymbol{\theta }_{(t-1)Q^{\prime }(1),}w_{1})+p_{t21}^{B}(\theta _{t2}|%
\boldsymbol{\theta }_{tQ(2),},\boldsymbol{\theta }_{(t-1)Q^{\prime
}(2),}w_{1})\right] ^{-1}
\end{equation*}%
In this case we have a default causal algebra for this type of
intervention where $\mathcal{G}$ remains invariant and probabilities change.
Clearly this algebra requires $D$ to provide additional information about
the extent certain probabilities will change. This in turn will be a
function of their domain knowledge. Otherwise, the algebra is a hybrid of
the ``do" algebra of the CEG adjusted to the RDCEG \cite{ThwaitesetalCause} on the deepest layer and Pearl's ``do" algebra \cite{Pearl2000} applied to the intermediate layer of the hybrid graph of the
plot.

\item $d_{w}\in \mathbb{D}_{W}.$ When defender $D$ connects with $A$'s agent
and tells them they are aware of the plot. This could have one of two
effects. The meeting might convince the agent to abort their plans and to
lie low for a while: in our notation to increase the probabilities $%
q_{it}^{\prime }$ of returning to the benign state. Alternatively, the agent
could continue their plans, but attempt to do these more covertly. This would
change what $D$ could expect to subsequently observe of $A$'s progress, the
speed they performed the tasks necessary for each of the active phases in
which $A$ lie as well as their likely use of various tasks needed to pass
through a phase. So $D$ would need to review and adapt most of the factors
in $\mathcal{P}$ to construct a new set of factors $\mathcal{P}_{d_{W}}$.
\end{enumerate}

Note that there are no intelligent counters to a $d_{p}\in \mathbb{D}_{P}$.

\subsubsection{Intelligent Reactions When Discovering a Plot is Betrayed}

Let $K_{dt}^{\ast }$ denote the indicator variable that an intervention $%
d\in \mathbb{D}$ betrays to $A$ that $D$ is aware of $A$'s plot at time $t$.
Then many interventions $d \in \mathbb{D}\setminus \left\{d_{\emptyset }\right\}$ that are not
securely covert are likely to increase the probability of such betrayal as
compared with the idle system: i.e. there is at least one $d \in \mathbb{D}\setminus \left\{d_{\emptyset }\right\}$ instigated at a time $t_{0}$ such that 
\begin{equation*}
P(K_{dt_{0}}^{\ast }=1)>P(K_{d_{\emptyset }t_{0}}^{\ast }=1)
\end{equation*}%
Such a discovery by $A$ is likely to have a significant effect on their
behaviour. So in a formal sense we discuss above $K_{dt}^{\ast }$ is a
potential cause of $A$'s reactions and therefore $D$'s beliefs about the
progress of the plot. This feature therefore needs to appear \emph{%
explicitly} in any graphical representation of the causal processes behind
the plot.

The simplest way of adapting $\mathcal{G}^{\prime }$ into a causal graph $%
\mathcal{G}^{+}$ that explicitly represents this process is to double the
number of active phases of the idle plot. So, if for example we assume that
it will take $A$ a time step to react to such a realisation, we replace each
active phase $\left\{ w_{1},w_{2},\ldots ,w_{m}\right\} $ by $\left\{
w_{1}^{-},w_{2}^{-},\ldots ,w_{m}^{-},w_{1}^{\ast },w_{2}^{\ast },\ldots
,w_{m}^{\ast }\right\} $ in the underlying RDCEG model. The new vertices $%
\left\{ w_{i}^{-},w_{i}^{\ast }:i=1,2,\ldots ,m\right\} $ split into the two
active phases $\left\{ w_{i}:i=1,2,\ldots ,m\right\} $ in $\mathcal{G}$ and $%
\mathcal{G}$' to represent the same phases when, respectively, $A$ is
unaware or aware, that $D$ has discovered $A$'s plot.

\begin{figure}
    \centering
\begin{tikzpicture}[main/.style = {draw, circle}, node distance={20mm},line width=1pt,>=stealth] 
\node[main] (1) at (-2,0) {$w_1^-$}; 
\node[main] (2) at (0,1) {$w_2^-$}; 
\node[main] (21) at (2,-1) {$w_{2\star}^-$}; 
\node[main] (3) at (0,-1) {$w_3^-$};
\node[main] (31) at (2,1) {$w_{3\star}^-$};
\node[main] (4) at (4,0) {$w_4^-$};
\draw[->] (1) -- (2);
\draw[->] (1) -- (3);
\draw[->] (2) -- (31);
\draw[->] (3) -- (21);
\draw[->] (21) -- (4);
\draw[->] (3.8,-0.4) -- (2.45,-1.15);
\draw[->] (31) -- (4);

\node[main] (1p) at (-2,4) {$w_1^+$}; 
\node[main] (2p) at (0,3) {$w_2^+$}; 
\node[main] (21p) at (2,5) {$w_{2\star}^+$}; 
\node[main] (3p) at (0,5) {$w_3^+$};
\node[main] (31p) at (2,3) {$w_{3\star}^+$};
\node[main] (4p) at (4,4) {$w_4^+$};
\draw[->] (1p) -- (2p);
\draw[->] (1p) -- (3p);
\draw[->] (2p) -- (31p);
\draw[->] (3p) -- (21p);
\draw[->] (21p) -- (4p);
\draw[->] (3.8,-0.4) -- (2.45,-1.15);
\draw[->] (31p) -- (4p);
\draw[->] (3.8,4.45) -- (2.5,5.15);
\draw[->] (1) -- (1p);
\draw[->] (2) -- (2p);
\draw[->] (21) to [out=300,in=60,looseness=3] (21p);
\draw[->] (3) to [out=180,in=180,looseness=2] (3p) ;
\draw[->] (31) -- (31p);
\draw[->] (4) -- (4p);
\end{tikzpicture}
\caption{An RDCEG graph $\mathcal{G}^{W-}$ for a bombing plot.}
\label{fig:fig4}
\end{figure}
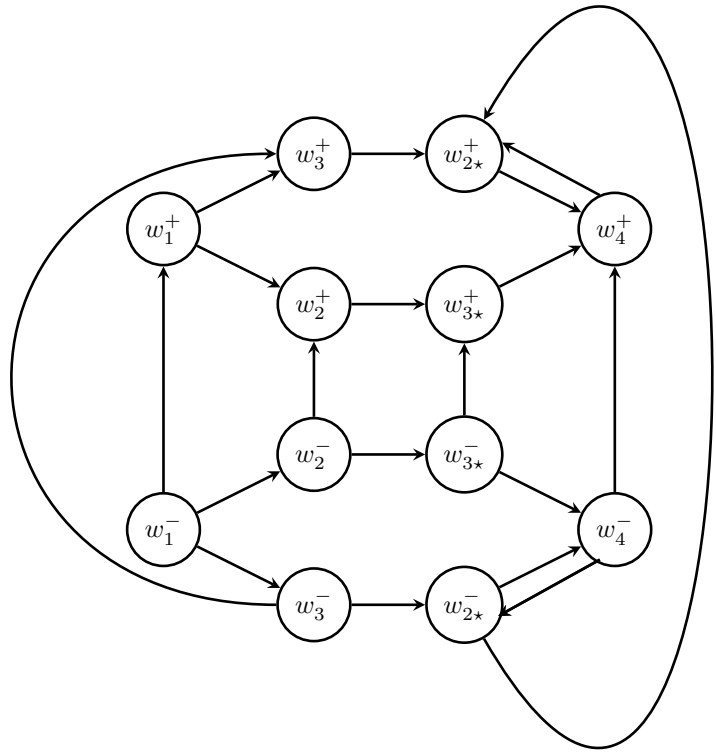

Then, under the assumption above, the semantics of this graph would have both a directed edge $\left( w_{i}^{-},w_{j}^{-}\right) $, $\left\{
w_{i}^{-},w_{j}^{-}:i,j=1,2,\ldots ,m\right\} $ and a directed edge $\left(
w_{i}^{\ast },w_{j}^{\ast }\right) $, $\left\{ w_{i}^{\ast },w_{j}^{\ast
}:i,j=1,2,\ldots ,m\right\} $ in $\mathcal{G}^{+W-}$, if and only if there is
a directed edge $\left( w_{i},w_{j}\right) $, $\left\{
w_{i},w_{j}:i,j=1,2,\ldots ,m\right\} $ in $\mathcal{G}^{W-}$ and a directed
edge $\left( w_{i}^{-},w_{j}^{\ast }\right) $ , $\left\{
w_{i}^{-},w_{j}^{\ast }:i,j=1,2,\ldots ,m\right\} $ in $\mathcal{G}^{+W-}$ if
and only if $i=j$. Note that once $A$ becomes aware of $D$'s discovery they
will continue to know this. So there can be no transitions from vertices in $%
\left\{ w_{i}^{\ast }:i=1,2,\ldots ,m\right\} $ to vertices in $\left\{
w_{j}^{-}, j=1,2,\ldots ,m\right\}$. So for example in our running example
the RDCEG of our new extended plot model is shown in Figure \ref{fig:fig4}.

The factors then need to be constructed that populate this new RDCEG.
Suppose the intervention will happen at time $t_{0}$. Then this involves $D$
specifying 
\begin{equation*}
\mathcal{P}_{\mathcal{G}^{+}\mathbb{D}}^{-W}=\left\{
p_{t_{0}d}^{-W},q_{tid}^{-\prime },q_{tid}^{-},p_{tkjd}^{-W}:1\leq i\leq
m,j\in E_{i},\text{ }k=0,1,2,\ldots ,K, \text{ } t=t_{0},t_{0}+1\ldots ,T, \text{ } d\in \mathbb{%
D}\right\}
\end{equation*}%
for the case when it is assumed that $A$ is ignorant that $D$ is aware their
plot, and
\begin{equation*}
\mathcal{P}_{\mathcal{G}^{+}\mathbb{D}}^{\ast W}=\left\{ p_{t_{0}d}^{\ast
W},q_{tid}^{\ast \prime },q_{tid}^{\ast },p_{tkjd}^{\ast W}:1\leq i\leq
m,j\in E_{i}, \text{ } k=0,1,2,\ldots ,K, \text{ } t=t_{0},t_{0}+1\ldots ,T, \text{ } d\in \mathbb{%
D}\right\}
\end{equation*}%
for the case when $A$ has become aware, plus the probability $\pi _{t_{0}d}$

\begin{equation*}
\mathcal{P}_{\mathcal{G}^{+}\mathbb{D}}^{"W}=\left\{ \pi _{t_{0}d},\text{ }d\in 
\mathbb{D}\right\}
\end{equation*}%
the common edge probability of the discovery being made. Note that the probability that $A$ discovers that $D$ is aware of their plot depends on $d$. All the meanings of
these probabilities in $\mathcal{P}_{\mathcal{G}^{+}\mathbb{D}}^{-W}$ and $%
\mathcal{P}_{\mathcal{G}^{+}\mathbb{D}}^{\ast W}$ are as they were in $%
\mathcal{P}_{\mathcal{G}^{\prime }\mathbb{D}}^{W}=\mathcal{P}_{\mathcal{G}%
\mathbb{D}}^{W}$ but now conditioned on the value of $K^{\ast }.$ The set of
factors associated with $\mathcal{P}_{\mathcal{G}^{+}}^{W}$ for potential
interventions $d\in \mathbb{D}$ at time $t_{0}$ is then 
\begin{equation*}
\mathcal{P}_{\mathcal{G}^{+}\mathbb{D}}^{W}\triangleq \mathcal{P}_{\mathcal{G%
}^{+}\mathbb{D}}^{-W}\cup \mathcal{P}_{\mathcal{G}^{+}\mathbb{D}}^{\ast
W}\cup \mathcal{P}_{\mathcal{G}^{+}\mathbb{D}}^{"W}
\end{equation*}

Note that such a discovery will not affect what $A$ needs to do to
accomplish their tasks to pass through the phases to successfully complete
their plot. Therefore, allowing for such a discovery will affect the
topology of $\mathcal{G}^{\prime B}\cup \mathcal{G}^{\prime S}.$ The
topology of these graphs are simply copied across the two different sets of
phases. Thus, more explicitly, to produce $\mathcal{G}^{+}$ we simply
concatenate%
\begin{equation*}
\mathcal{G}^{+B}\cup \mathcal{G}^{+S}\triangleq \mathcal{G}^{\prime -B}\cup 
\mathcal{G}^{\prime -S}\cup \mathcal{G}^{\prime \ast B}\cup \mathcal{G}%
^{\prime \ast S}
\end{equation*}%
on to $\mathcal{G}^{+W-}$ by associating both the relevant vertices labeled
by the two new sets of phases where the topology of both $\mathcal{G}%
^{\prime -B}\cup \mathcal{G}^{\prime -S}$ and $\mathcal{G}^{\prime \ast
B}\cup \mathcal{G}^{\prime \ast S}$ are equal to the topology of $\mathcal{G}%
^{\prime B}\cup \mathcal{G}^{\prime S}$. So then 
\begin{equation*}
\mathcal{G}^{+}=\mathcal{G}^{+W-}\cup \mathcal{G}^{+B}\cup \mathcal{G}^{+S}
\end{equation*}%
The probabilities embellishing such new graphs will then, of course, need to
be adapted to each new regime. Here - extending our notation in an obvious
way - we need to elicit the conditional probabilities in

\begin{equation*}
\mathcal{P}_{\mathcal{G}^{+}}^{B}=\mathcal{P}_{\mathcal{G}^{+}}^{-B}\cup 
\mathcal{P}_{\mathcal{G}^{+}}^{\ast B}\text{ and }\mathcal{P}_{\mathcal{G}%
^{+}}^{S}=\mathcal{P}_{\mathcal{G}^{+}}^{-S}\cup \mathcal{P}_{\mathcal{G}%
^{+}}^{\ast S}
\end{equation*}%
where $\mathcal{P}_{\mathcal{G}^{+}}^{-B}$ and $\mathcal{P}_{\mathcal{G}%
^{+}}^{-S}$ correspond to factors when $A$ is unaware, and $\mathcal{P}_{%
\mathcal{G}^{+}}^{\ast B}$ and $\mathcal{P}_{\mathcal{G}^{+}}^{\ast S}$ when 
$A$ is aware.

Extra elicitations efforts will be required by $D$ to deliver these additional factors. However, commonly this additional effort is
not substantial. For example, suppose $D$ believed that $A$ were timid -
meaning so that if they learned the plot has been discovered at time $t_{0}$
then $A$ would abort the current plot and transition to the inactive state $%
w_{0}$. Then, in the notation above, we would simply set the probabilities $%
q_{tj}^{\prime }=1,$ $j=1,2,\ldots ,m$, $t=t_{0},t_{0}+1,\ldots ,T$, and
all other probabilities associated with $K^{\ast }=1$ would be
irrelevant. In effect $D$ will have then implicitly and simultaneously enacted a 
$d\in \mathbb{D}_{P}$. In our notation none of the other probabilities in $%
\mathcal{P}_{\mathcal{G}^{+}}^{\ast B}\cup \mathcal{P}_{\mathcal{G}%
^{+}}^{\ast S}$ would then need to be elicited.

On the other hand, if on learning $K^{+}$ they would either be commanded by $%
A $ to abort or not - in which case the agent would carry on regardless -
then again most of the probabilities associated after this discovery would
be simple functions of probabilities populating the idle system. This latter
case would necessarily hold for example if $D$ believed that $A$'s agent
were not agile and was incapable or unwilling to change their MO in the
light of this information.

So in practice we find the implications of this adaptation is not large.
Even when this is not so it is often reasonable for $D$ to believe that most
of these conditional probabilities will duplicate themselves over the new
classes. 

Above we have considered predictions only associated with interventions for
which discovery is binary. However, there will be other settings where whilst a $d\in \mathbb{D}$ does not fully disclose $D$'s awareness, it is
possible for $A$ to learn \emph{something} about that intervention. This makes
the modelling somewhat more complex since to build a causal model for such
scenarios, $D$ would need to model with further discovery states expressed
through $\boldsymbol{K}_{d}$. The principles illustrated above though still
apply - this added complexity just adds more states to the underlying plot.

\subsection{Extension of Example for an Intervention with an Intelligent Adversary}
We return to our illustration of Example 1 as presented in Section \ref{subsec:example}. 

Specifically, rather than filtering all the data as we did in that section, we imagine considering making a blocking intervention, in this case confiscating $A$'s agent's passport for a purportedly unrelated crime, after week 2024-01-08. 

To understand the implications of such an intervention, we may only use the observed (synthetic) data until this time point. We can then explore the effects of predicting the probabilities of $A$'s progress through the plot from this time point forwards. We can do this for the baseline case where no intervention is made, and compare these probabilities to their equivalents in the case that such an intervention is made. In the case of the intervention, we alter the model dynamics as described in Section \ref{subsubsec:intelligent}, to account both for the intervention itself and also the $A$'s agent's intelligent response to the intervention.     

\begin{figure}
\includegraphics[width=\textwidth]{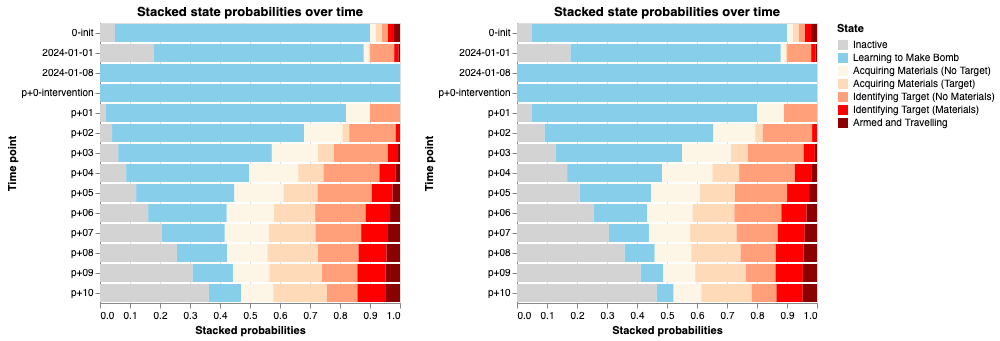}
\caption{Stacked bar-charts of the evolution of predicted probability estimates of the plot phase of the adversary $A$, comparing the idle model (left panel) with the intervened model, where $D$ confiscates the passport of $A$'s agent, allowing for an intelligent reaction (right panel).\label{fig:intervene_visual}
}
\centering
\end{figure}

\begin{figure}
\includegraphics[width=\textwidth]{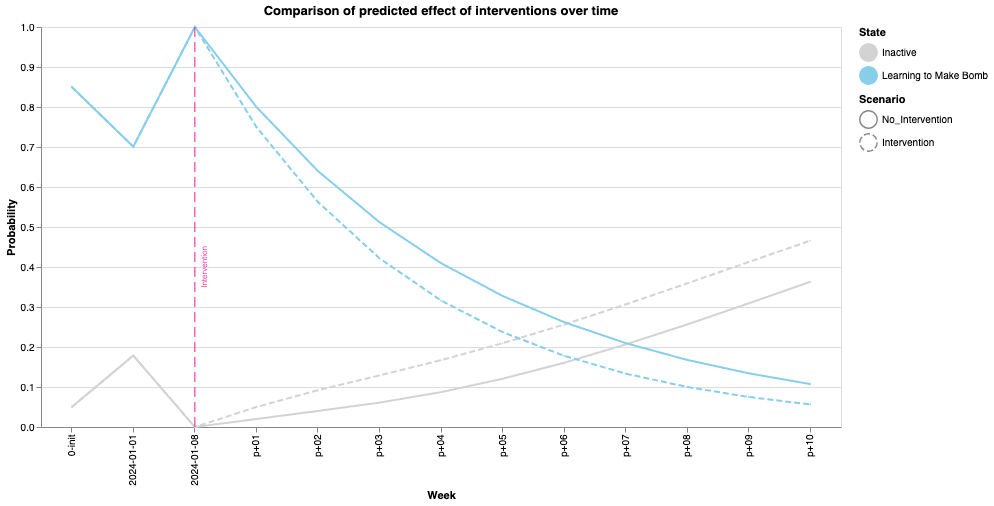}
\caption{Line charts of the state probabilities over time, for the \emph{Inactive} (grey) and \emph{Learning to Make Bomb} (blue) plot phases. The solid line shows the baseline predictions, with the dashed line showing the predictions in the case of the proposed intervention.\label{fig:intervene_compare}
}
\centering
\end{figure}

Figure \ref{fig:intervene_visual} compares the 
predicted evolution of the estimates of the plot-phase probabilities, in the idle model (left panel) and the intervened model (right panel). Although there is no intervention in the idle model in the left panel, we create a \emph{null} intervention by duplicating the probabilities from the previous time point, to ensure the charts are aligned to aid comparison. The charts show that the effects of this intervention are relatively minimal, but at each time point after the intervention there is an increased probability that the adversary becomes \emph{Inactive} (largely attributable to them abandoning the plot). This comes at the expense of remaining in the \emph{Learning to Make Bomb} phase. The evolution of the small differences in the probabilities for these two states for the baseline and intervened cases can be seen in Figure \ref{fig:intervene_compare}.

\section{Discussion}

We have demonstrated how a graphical causal analysis can be applied to
plot models in a way that can form part the support tools for both strategic
analyses and real time decision support. By following ARA protocols and an
appropriate customised causal graph we can adapt causal analyses so that these apply to study a plot. We show here that we simply need to embellish a graph
of the idle system to include within $D$'s description of the plot features that describe $D$'s belief about $A$'s awareness they have been discovered. We achieve this through adding more positions at the deepest level of the hierarchy and a description of $A$'s capability to deviate from their routine ways of
progressing their plot, through, if necessary, embellishing the task vector
that describe how $A$ might circumvent $D$'s attempts to frustrate them.
Then, in the notation above, using this embellished graph $\mathcal{G}^{+}$
instead of the graph $\mathcal{G}$ of the idle system, $D$ simply needs to
specify how certain factors in their joint density of what might happen
given various entertained interventions $d\in \mathbb{D}$ are enacted. This
then provides all inputs needed to calculate their SEU decisions. This
protocol depends on $D$'s belief that their chosen graphical model $%
\mathcal{G}^{+}$ is causal: not only in the sense that they believe it is invariant to their choice of intervention $d\in \mathbb{D}$, but that $A$
also believes that $\mathcal{G}^{+}$ faithfully describes the underlying
process. In general we discuss in \cite{Ramiahetal23} why the former is
predicated on the latter. However, in the case of plots this assumption is
usually an extremely plausible one to make.

Although, for brevity, we have illustrated the causal analysis of one type
of plot associated with monitoring a terrorist attack as in \cite{BuuninandSmith19,ShneivB&S21}, there are now many other types of plot
this causal technology is now being transferred to. The basic protocol we
describe above has been successfully implemented in a number of such cases.
Of course, the adoption of these modelling methodologies is still in its
infancy. We are currently examining how standard causal discovery algorithms
could be adapted so as to learn about plots. It appears that this is
possible through the translation of automatically learned graphs of the idle
system and the translation of estimated factors from the idle system to the
controlled one. In this new domain it is becoming clear that - unlike in
more conventional settings - the embedding of probabilistic expert
judgements are a vital component of this discovery process. So a Bayesian SEU
approach in conjunction with suitable model selection algorithms seems to be
the most promising way forward here.

However, what we hope to have demonstrated here is that because plot models
are graphical, they can provide a very powerful framework around which a
defender can formally and transparently explore the efficacy of the
potential interventions that might be available to them. Moreover, we demonstrated that we can develop libraries of such plots \citep{Drury}. In particular ARA
technologies when combined with causal graphical inferential frameworks can
sometimes make the complex task of probabilistic models in adversarial
domains much more feasible and transparent than they would otherwise be.

\bibliographystyle{unsrtnat}
\bibliography{biblio}

\end{document}